\documentclass{imsart}

\usepackage{amscd,amsfonts,amsmath,amssymb,amsthm,bbm,bm,latexsym,mathrsfs}
\usepackage{epsfig,graphics,natbib,graphicx,subfigure,longtable}
\usepackage[english]{babel}
\usepackage[usenames]{color}
\usepackage{booktabs}
\usepackage{sgame, float}
\usepackage[top=1.7in, bottom=1.7in, left=0.9in, right=0.9in]{geometry}
\usepackage{epstopdf}
\usepackage{algorithmicx}
\usepackage{algpseudocode}
\usepackage{algorithm}
\usepackage{rotating}
\usepackage{natbib}

\allowdisplaybreaks
\makeatother

\newfloat{algorithm}{tbp}{loa}
\providecommand{\algorithmname}{Algorithm}
\floatname{algorithm}{\protect\algorithmname}

\startlocaldefs
\endlocaldefs

\begin{document}

\begin{frontmatter}

\title{Advances in Bayesian random partition models: A comprehensive review}
\runtitle{Bayesian clustering}


\author{\fnms{Clara} \snm{Grazian}\ead[label=e1]{clara.grazian@sydney.edu.au}}
\address{\printead{e1}}
\affiliation{University of Sydney}

\runauthor{Grazian, C.}

\begin{abstract}
Clustering is a crucial task in various domains of knowledge, including medicine, epidemiology, genomics, environmental science, economics, and visual sciences, among others. Methodologies for inferring the number of clusters have often been shown to be inconsistent, and incorporating a dependence structure among clusters introduces additional challenges in the estimation process. In a Bayesian framework, clustering is performed by treating the unknown partition as a random object and defining a prior distribution for it. This prior distribution can be induced by models assumed for the observations or directly defined on the partition itself. However, recent findings have revealed difficulties in consistently estimating the number of clusters and, consequently, the partition. Furthermore, summarizing the posterior distribution of the partition remains an open problem due to the high dimensionality of the partition space. This study aims to review Bayesian approaches for random partition models, highlighting the advantages and disadvantages of each method, and suggesting potential avenues for future research. 
\end{abstract}



\end{frontmatter}

\section{Introduction}

An important task in statistical modeling is the identification of groups or partitions among observations, aiming to group together those that exhibit greater homogeneity in a specific aspect compared to other clusters. Clustering serves as an initial step in numerous analyses. For instance, in regression, it is common to utilize homogeneous groups to explore associations with particular covariates. However, while clustering is a vital step, it is also a delicate task as the interpretation of the resulting groups is subjective.

There are two main approaches to clustering: distance-based clustering methods, such as $k$-means and hierarchical clustering, which define similarity among observations based on a chosen distance metric; and model-based clustering approaches, which assume a probabilistic model and probabilistically assign observations to different clusters. For model-based clustering with a fixed and known number of groups, mixture models are commonly used. However, a drawback of using these models is the need to estimate or select the number of clusters in advance. Model selection techniques, such as the deviance information criterion (DIC) \citep{celeux2006deviance} or the integrated likelihood criterion (ICL) \citep{biernacki2000assessing}, are available. Nevertheless, the performance of each method can vary depending on the specific problem, and different criteria may disagree regarding the true number of components in the underlying model.

Alternatively, it is possible to consider the number of components as an unknown parameter and define a prior distribution for it \citep{nobile2004posterior}. In this context, the prior distributions used for the number of components or the component parameters can significantly influence posterior estimation. For instance, studies by \cite{richardson1997bayesian} and \cite{jasra2005} demonstrate that a Gaussian mixture model with a prior distribution having a large variance on the component means tends to favor smaller values for the posterior distribution of the number of components (and consequently, the number of clusters). Furthermore, research indicates that the posterior distribution on the number of components can diverge when there is misspecification of the component distributions, as shown in studies such as \cite{woo2006robust}, \cite{woo2007robust}, \cite{rodriguez2014univariate}, and \cite{cai2021finite}. In this work, it is assumed that the component distributions are correctly specified, and we focus on priors for the number of components. 

The nonparametric extensions of finite mixture models, which allow for an infinite number of components, often rely on Dirichlet processes (DP) \citep{Ferguson1973}. Dirichlet processes have a significant role in Bayesian nonparametrics, not only for clustering but also for density estimation, due to their computationally manageable representations. Notably, the stick-breaking representation \citep{sethuraman:stick}, the P\'olya urn representation \citep{blackwell1973ferguson}, and the Chinese restaurant process \citep{aldous1985exchangeability} are frequently used. One key characteristic of the Dirichlet process is its ability to assign probability one to a set of countable, discrete distributions. While this property poses a limitation for density estimation, which is usually overcome through the definition of Dirichlet process mixture models, it proves useful in clustering as it automatically groups observations.

Although consistency in $L_1$ to the true density has been demonstrated for Dirichlet process mixtures in density estimation, achieving the minimax optimal rate up to logarithmic factors \citep{ghosal1999posterior, ghosal2001entropies, lijoi2005consistency, tokdar2006posterior, ghosal2007convergence, walker2007rates, kruijer2010adaptive, wu2010l1, nguyen2013convergence}, these results cannot be readily extended to study consistency for the number of clusters. This is because any mixture with $k$ components can be approximated in $L_1$ by another mixture with $k+1$ components (or generally $k'>k$). Recent work by \cite{miller2014inconsistency} analytically proves the inconsistency of the posterior distribution on the number of components for a broad class of infinite mixtures, including DP mixture models with various forms of component distributions.

A particular mention should be made regarding clustering in the presence of covariates. This refers to the partitioning of a set of experimental units, where the probability of each partition depends on the covariates. In other words, observations with similar or equal levels of covariates are more likely to be clustered together. In model-based clustering, the dependence on covariates can be incorporated into the cluster probabilities. Examples of such approaches can be found in works by \cite{pawlowsky1992spatial}, \cite{fernandez2002modelling}, \cite{tjelmeland2003bayesian}, \cite{neelon2014multivariate}, and \cite{paci2018dynamic}. To address biases arising from the sum-to-one constraint of the probability vector, \cite{mastrantonio2019new} propose a logit-Gaussian process. For a recent review of dependent Dirichlet processes, we refer to \cite{quintana2020dependent}, and for a comprehensive review of Bayesian clustering methodologies, we also refer the reader to \cite{wade2023bayesian}, focused on Bayesian covariate-dependent mixture models.

This study aims to examine the advantages and disadvantages of various approaches based on random partition models found in the literature, in order to provide insights for future research directions.

The paper is organized as follows: Section \ref{sec:rpm} provides the definitions of random partition models, along with the notation used throughout the paper. Section \ref{sec:InducedRpm} explores induced random partition models based on both finite and infinite mixture models. Section \ref{sec:ppm} discusses product partition models and other prior distributions for partitions. Section \ref{sec:clust_subpop} addresses the problem of clustering populations, while Section \ref{sec:posteriors} introduces approaches to estimate the optimal partition once its posterior distribution is available. Finally, Section \ref{sec:conclu} presents the concluding remarks of the paper.

\section{Random partition models}
\label{sec:rpm}

Model-based clustering involves randomly allocating observations to clusters identified by the model. Let $[n] = \{1, \ldots, n\}$ be a set of $n$ indices, and define $\rho_n = (S_1, \ldots, S_K)$ as a random partition of the set $[n]$, where $K = |\rho_n| \leq n$ represents the number of non-empty and mutually exclusive subsets. The sets $S_h$ satisfy $\cup_{S_h \in \rho_n} S_h = [n]$ and $S_{\ell} \cap S_h = \emptyset$ for $\ell \neq h$. The random partition $\rho_n \in \mathcal{P}_n$, where $\mathcal{P}_n$ denotes the set of all possible partitions of $[n]$, whose size is the $n$-th Bell number. The size of $\mathcal{P}_n$ rapidly increases with $n$, making analytical computations infeasible, even for small sample sizes. 

It is common to represent a partition $\rho_n = (S_1, \ldots, S_{K})$ using class label memberships $C_1, \ldots, C_n$, where $C_i \in [K]$ and
$$
C_i = h \Leftrightarrow i \in S_h \qquad \mbox{for } i \in [n] \mbox{ and } 1 \leq h \leq K.
$$

A random partition $\rho_n$ is said to be exchangeable if its distribution is invariant under permutations of $[n]$. The distribution of the partition $\rho_n$ is called the exchangeable partition probability function (EPPF) \citep{pitman1995exchangeable}, expressed as
$$
P(\rho_n = (S_1, \ldots, S_{K})) = p(|S_1|, \ldots, |S_{K}|) = p(n_1, \ldots, n_K),
$$
which is a function $p: \mathbb{N}^* = \cup_{h=1}^\infty \mathbb{N}^h \rightarrow [0,1]$ symmetric in its argument and invariant under permutation of the elements of $[n]$, and where $\mathbb{N}$ represents the set of natural numbers. It is worth noting that many works, including this one, use this definition in terms of aggregated probabilities, i.e., probabilities for cluster sizes $\sum_{\rho_n \in \mathcal{A}} P(\rho_n = {S_1, \ldots, S_K})$, where $\mathcal{A} = \{\rho_n \in \mathcal{P}^n: \rho_n \text{ has cluster sizes } (n_1, \ldots, n_K)\}$. For an interesting discussion on definitions of the probability mass function of each partition, refer to \cite{lee2022rich}.

The EPPF has the following properties. Let $\mathbf{n}$ be the infinite sequence $(n_1, n_2, \ldots, n_K, 0, 0, \ldots)$. Then, $p(1) = 1$ and 
\begin{align*}
p(\mathbf{n}) = \sum_{h=1}^{K(\mathbf{n)}+1} p(\mathbf{n}^{h+}) \quad \forall \; \mathbf{n} \in \mathbb{N}^*.
\end{align*}
Here, $\mathbf{n}^{h+}$ corresponds to $\mathbf{n}$ with the $h$-th element increased by one unit, and $K(\mathbf{n})$ represents the number of non-zero components of $\mathbf{n}$. Moreover, a notable property of the EPPF is sample size consistency \citep{de2013gibbs}: $p(\rho_n)$ can be derived from $p(\rho_{n+1})$ by marginalizing the last element. 

It is important to note the distinction between the number of components and the number of clusters. For a given $K$, $K(\mathbf{n}) = K_+$ is defined as the number of components that generated the data, i.e., $K_+ = \sum_{h=1}^K \mathbb{I}\{n_h > 0\}$, where $n_h = \#\{i: C_i = h\}$ counts the observations allocated to component $h$. For further discussion, see \cite{argiento2019infinity}. 

The EPPF is associated with the prediction probability function (PPF). Consider an exchageable sequence $(X_1, X_2, \ldots)$ of random variables and the distribution of the first $n$ variables. Let $X_i^*$, $j=1, \ldots, K$, define the $K \leq n$ unique values among $(X_1, \ldots, X_n)$. The posterior predictive distribution of the $(n+1)$-th observation can be written as 
$$
X_{n+1} | X_1, \ldots, X_n \sim \sum_{i=1}^K p_i(\mathbf{n}) \delta_{X_i^*} + p_{K+1}(\mathbf{n})
$$
with weights $p_i(\mathbf{n})$. This is known as predictive probability function (PPF), and it is given by
$$
p_j(\mathbf{n}) = \frac{p(\mathbf{n}^{j+})}{p(\mathbf{n})} \quad 1 \leq j \leq K + 1.
$$
While the definition of a PPF directly derives from an EPPF, the converse is not necessarily true. \cite{lee2013defining} provide a necessary and sufficient condition for arbitrary PPFs to define an EPPF.

In the following sections, we review methods to define partition models for $\rho_n$. Some of these methods result in an analytical form of the EPPF, while others are not associated with a closed-form expression of the EPPF.

\section{Induced random partitions models}
\label{sec:InducedRpm}

Consider continuous random variables $Y_1, \ldots, Y_n$. A possible way to define a random partition probability distribution is by constructing a hierarchical model on the observations and inducing a model on the random partition. For example, a hierarchical model of this type can be expressed as follows:
\begin{align}
Y_1, \ldots, Y_n &\sim g(y_1, \ldots, y_n | \theta_1, \ldots, \theta_n) \nonumber \\
\theta_1, \ldots, \theta_n | F & \sim F \label{eq:partmod1}\\
F &= \mbox{discrete RPM}, \nonumber
\end{align}
\noindent where $RPM$ denotes a random probability measure. The discreteness of $F$ implies the presence of ties among the realisations of the random vector $(\theta_1, \ldots, \theta_n)$. Let $\theta^*_1, \ldots, \theta^*_{K}$ denote the unique values of $\theta_1, \ldots, \theta_n$. The partition $\rho_n$ can be redefined as follows: $
C_i = h \Leftrightarrow \theta_i = \theta_h^*$. 
Thus, $S_h = \{i \in [n]: \theta_i = \theta_h^*\}$. Alternatively, the $h$-th unique value $\theta^*_h$ can also be denoted as $\theta_{C_i}^*$.
A common choice for inducing a partition on the observations is to use mixture models.

\subsection{Finite mixture models}
\label{sub:fmm}

Consider a model for independent observations, such that $g(y_1, \ldots, y_n | \theta_1, \ldots, \theta_n) = \prod_{i=1}^n g(y_i | \theta_i)$ in Equation \eqref{eq:partmod1}. If the random $F$ is discrete with $K$ atoms $\theta_1^*, \ldots, \theta_K^*$, then model \eqref{eq:partmod1} reduces to a finite mixture model. Generally, a finite mixture model for observation $Y_i$ \citep{fruhwirth2006finite, fruhwirth2019handbook} is given by:
 \begin{equation}
\label{eq:finitemix}
g(y_i | \pi_1, \ldots, \pi_K, \theta^*_1, \ldots, \theta^*_K) = \sum_{h=1}^{K} \pi_h f_h(y_i | \theta^*_h) \qquad i=1, \ldots, n.
\end{equation}
Here, $(\pi_1, \ldots, \pi_{K})$ are weights satisfying $\pi_h \geq 0$ for $h=1, \ldots, K$ and $\sum_{h=1}^{K} \pi_h = 1$. The term $f_h(\cdot | \theta^*_h)$ represents a probability distribution indexed by component-specific parameters. Typically, the component distributions are assumed to be from the same family, so we have $f_h(\cdot | \theta^*_h) = f(\cdot | \theta^*_h)$.  


\subsubsection{Overfitted mixtures.} In practical examples, it is common to employ sparse mixture models where a fixed, overfitting value of $K$ is chosen \citep{rousseau2011asymptotic}, along with a symmetric Dirichlet prior distribution for the weights with a small parameter $\gamma$. In this approach, although $K$ is fixed, the number of clusters $K_+$ is a random variable because some components will have weight $\pi_j = 0$ or some component might be merged. Therefore, the number of clusters is identified as $K_+ = \sum_{h=1}^K \mathbb{I}\{\pi_h > 0\}$. \cite{grazian2018jeffreys} investigate the properties of Jeffreys' prior distributions in this context and demonstrate consistent estimation of the number of clusters. The prior distribution induced on the random partition by a sparse mixture model approaches the Ewens distribution when $\gamma  = \alpha/K \rightarrow 0$, where $\alpha$ is a constant and can be seen as the concentration parameter of the corresponding Dirichlet process (see Section \ref{sub:dpmixtures}). The Ewens distribution is the distribution induced on the partition by a Dirichlet processThe terminology follows Fr{"u}hwirth-Schnatter et~al. (2021) and McCullagh and Yang (2008). Consequently, as we will see for the Dirichlet process, the estimation process for a sparse mixture model tends to concentrate on a large number of small clusters as $n$ increases.

\cite{grazian2018jeffreys} propose overfitted mixtures to monitor IP packages for software deployment, in particular via flow-entry retrieval.

\subsubsection{Static mixture models.} Alternatively, it is possible to consider a random number of components $K \sim p_K(k)$, where $p_K$ is a probability mass function on $\mathbb{N}$ such that $\sum_{h=1}^\infty p_K(h) = 1$ and $p_K(h) > 0$ for $\forall h$ \citep{nobile1994bayesian}. Including a prior distribution $p_K(k)$ has the effect that both $K_+$ and $K$ are random a priori. \cite{kruijer2010adaptive} and \cite{nobile1994bayesian} prove consistency for the number of components when the component distributions are correctly specified. The assumption of correct specification of the component distribution is quite strong, as using Gaussian components usually only approximates the true model of the observations. However, this approximation can lead to the estimation of an increasing number of components as $n$ increases. 

A finite mixture model with a prior distribution on the number of components, 
induces a valid EPPF that is available in closed form \citep{green2001modelling, mccullagh2008many, miller2018mixture}:
\begin{equation}
p(\rho_n = (S_1, \ldots, S_{K_+})) = \sum_{\ell=1}^{\infty} \frac{\ell_{(K_+)}}{(\gamma \ell)^{(n)}} p_K(\ell) \prod_{s \in (S_1, \ldots, S_{\ell})} \gamma^{(|s|)}
\label{eq:EPPF_FM}
\end{equation}
where $b^{(m)} = b (b+1) \ldots (b+m-1)$ and $b_{(m)} = b (b-1) \ldots (b-m+1)$, $b^{(0)}=1$ and $b_{(0)}=1$, and where it is assumed that $(\pi_1, \ldots, \pi_{K})$ follows a Dirichlet prior distribution $Dir(\gamma, \ldots, \gamma)$. This model, with a fixed value of $\gamma$, which does not depend on the number of components, is called static mixture model. 

Equation \eqref{eq:EPPF_FM} reveals that the EPPF of a finite mixture model is a symmetric function of the cluster size, and the distribution of $\rho_n$ is invariant under permutations of $[n]$. Using Equation \eqref{eq:EPPF_FM}, we can derive the distribution of the number of components $K$ conditional on the number of clusters $K_+$:
$$
p(K=k | K_+ = k_+) = \frac{1}{\sum_{\ell=1}^{\infty} \frac{\ell_{(k_+)}}{(\gamma \ell)^{(n)}} p_K(\ell)} \frac{k_{(k_+)}}{(\gamma k)^{(n)}} p_K(k)
$$
and the distribution of the number of clusters conditional on the number of components:
$$
p(K_+ = k_+|K=k) = \frac{k_{(k_+)}}{(\gamma k)^{(n)}} \sum_{\mathbf{S}: |\mathbf{S}|=k_+ } \prod_{s \in (S_1, \ldots, S_{k_+})} \gamma^{(|s|)}.
$$
where $\mathbf{S}=\{S_h: |S_h|>0\}$. Finally, the conditional EPPF of a static mixture model is given by:
$$
p(|S_1|, \ldots, |S_k| | K_+ = k) = \frac{1}{Const_{k}} \prod_{h=1}^{k} \frac{\Gamma(n_h + \gamma)}{\Gamma(n_h + 1)}
$$
where $Const_k$ is the normalizing constant obtained by summing over all labeled cluster sizes whose sum is equal to $n$. As expected, this conditional EPPF depends on $\gamma$, and for $\gamma=1$, it represents the uniform distribution over all partitions.

Equation \eqref{eq:EPPF_FM} is valid when choosing a symmetric Dirichlet prior distribution on $(\pi_1, \ldots, \pi_{K})$. The choice of the value of $\gamma$ influences the entropy of the vector of weights: small values of $\gamma$ are associated to a low entropy while large values of $\gamma$ are associated to large entropy in $(\pi_1, \ldots, \pi_{K})$. In case of $\gamma = 1$, \cite{gnedin2010species} derives a form for $p_K(k)$ and \cite{stephens2000bayesian} and \cite{nobile2004posterior} propose $p_K(k)$ to be Poisson. In particular, when $p_K(k) = Pois(k-1 | \lambda)$ and $\gamma=1$, the finite mixture model has a stick-breaking representation \citep{argiento2019infinity}. \cite{richardson1997bayesian} and \cite{miller2018mixture} use $\gamma=1$ with a uniform prior $p_K(k)$ over $\{1, 2, \ldots, K_{\max}\}$. Sparse finite mixtures can be considered a special case of finite mixture models with an unknown number of components, because $p_K(k) = \mathbb{I}\{k \leq K_{max}\}$ puts all prior mass on a fixed number of components $K_{max}$. 

The use of $\gamma=1$ can introduce bias in the estimation of the number of clusters. According to \cite{fruhwirth2021generalized}, when $\gamma=1$, the expected value of $K_+$ tends to be close to the expected value of $K$ for most of the available prior distributions $p_K(k)$ in the literature (such as Poisson, uniform, geometric, and beta-negative-binomial). However, \cite{grazian2020loss} provide a decision-theoretic justification for using a beta-negative-binomial distribution with parameters $(1, \alpha,\beta)$. The choice of parameters allows control over prior expectation and variance, but it is recommended to use a value of $\gamma$ smaller than one, such as $\gamma = \frac{1}{2}$.

Furthermore, \cite{gnedin2006exchangeable} demonstrate that a finite mixture models, with a Dirichlet prior on the weights, fixed hyperparameter $\gamma$ and an unknown number of components, is equivalent to a mixture model with an infinite number of components and a Gibbs-type prior on the random partition. This model is the only finite mixture that induces a Gibbs-type prior (see Section \ref{sub:dpmixtures}). Interestingly, in this case, as $n \rightarrow \infty$, the number of clusters $K_+$ behaves similarly to the number of components $K$:
$$
|p(K_+=k_+ | y_1, \ldots, y_n) - p(K=k | y_1, \ldots, y_n)| \rightarrow 0 \qquad n \rightarrow \infty.
$$

\subsubsection{Dynamic mixture models.} 
Assuming the same $\gamma$ for all $K$ is a specific modeling choice that simplifies the implementation of known algorithms. To extend the static finite mixture model with a constant $\gamma$, \cite{mccullagh2008many} introduce the dynamic finite mixture model, where $\gamma_K = \alpha / K$. This means that the parameters of the Dirichlet distribution for the weights of the finite mixture model decrease as the number of components increases. The dynamic model reduces the impact of the experimenter's choice of $\gamma$. With increasing $K$, the symmetric Dirichlet distribution for the mixture weights becomes more concentrated around the boundary of the simplex, resulting in a more conservative estimation of $K_+$ and allowing the distribution of $K_+$ to differ from the distribution of $K$. Specifically, as $\gamma_K$ increases, the prior variance of the mixture weights decreases, leading to more balanced weights. Conversely, as $\gamma_K$ decreases (with a larger number of components), the prior variance increases, favoring more unbalanced weights. 

For a dynamic mixture model, the EPPF can be expressed as:
$$
p(\rho_n = (S_1, \ldots, S_{K_+})) = p_{DP}(\rho_n = (S_1, \ldots, S_{K_+})) \times \sum_{K = K_+}^{\infty} p_K(K) R_{K_+}^{K, \alpha}
$$
where $p_{DP}(S_1, \ldots, S_{K_+})$ is the probability mass function of the Ewens distribution (which will be defined in Section \ref{sub:dpmixtures}), and
$$
R_{K_+}^{K, \alpha} = \prod_{h=1}^{K_+} \frac{\Gamma(n_h + \frac{\alpha}{K})(K-h+1)}{\Gamma(1+\frac{\alpha}{K}) \Gamma(n_h) K}.
$$
The dynamic finite mixture model can be seen as a natural generalization of the Dirichlet process mixture model but does not fall into the class of Gibbs-type priors. Dynamic mixture models are characterized by a slower decrease in the difference between $\mathbb{E}[K_+]$ and $\mathbb{E}[K]$ as $\alpha$ increases compared to the static finite mixture model. This allows for larger differences even for large values of $\alpha$ because $\gamma$ decreases as $K$ increases, preventing $K_+$ from increasing too quickly.

The conditional EPPF of a dynamic mixture model can be expressed as:
$$
p(\rho_n = (S_1, \ldots, S_{k}) | K_+ = k) = p(|S_1|, \ldots, |S_{k}| | K_+ = k) = \frac{\sum_{\ell=1}^{\infty} p_K(\ell) \frac{b_{\ell, k}}{\Gamma\left(\frac{\alpha}{\ell}\right)^k} \prod_{h=1}^k \frac{\Gamma(n_h + \frac{\alpha}{\ell})}{\Gamma(n_h + 1)}}{\sum_{\ell=1}^{\infty}p_K(\ell) \frac{b_{\ell,k}}{\Gamma\left(\frac{\alpha}{\ell}\right)^k} Const_k}
$$
where $b_{\ell, k}$ is a constant depending on $k$, and $Const_k$ is the normalizing constant. Unlike the conditional EPPF of the static mixture model, this formula also depends on $\alpha$ and the prior distribution on the number of components, $p_K(\ell)$. This demonstrates that the dynamic mixture model offers more flexibility in defining the prior distribution on the number of clusters. However, the dependence on $p_K(\ell)$ implies that the choice of the prior distribution has a stronger impact on the induced prior distribution on the partitions.

It can be noticed that, keeping $\gamma$ fixed, for $K \rightarrow \infty$, the prior on the random distribution $F$ converges to a DP with scale parameter $\gamma$ and, therefore, has relationships with the case where $K$ is left random. For example, \cite{petrone1999bayesian} proposes an approach of this type for density estimation when data is defined in a closed interval. For a dynamic mixture model, this means that the sum $\sum_{j=1}^K \gamma/K = \gamma$ remains constant as $K$ varies, while for a static mixture model $\sum_{j=1}^K \gamma = K \gamma$, which increases with $K$; then. It is evident that the two models have different effects on the induced random partition.

Static and dynamic mixture models are used in many areas of science. For example, \cite{miller2018mixture} and \cite{grazian2020loss} analyze leukemia subtypes from gene expression in different patients. Identifying cancer types is crucial as it enables patient-specific treatments. Additionally, applications in other fields such as proteomics \citep{grazian2020loss}, social science \citep{fruhwirth2021generalized}, and clinical testing \citep{fruhwirth2021generalized} have also appeared, for both univariate and multivariate mixture models.

\subsubsection{Computational aspects.} 
In terms of computation, \cite{richardson1997bayesian} introduce reversible jump Markov Chain Monte Carlo (RJMCMC) for mixtures with univariate Gaussian components and a fixed parameter $\gamma$ that does not vary with $K$. \cite{miller2018mixture} generalize the Chinese restaurant process (CRP) sampler of \cite{Jain2004} to the case of finite mixtures, where the number of components $K$ is inferred, and the number of clusters $K_+$ is derived through post-processing. More recently, \cite{fruhwirth2021generalized} introduce a telescoping algorithm, an MCMC algorithm for mixtures that updates the number of clusters $K_+$ and the number of components $K$ simultaneously during the sampler without resorting to RJMCMC. The telescoping algorithm is implemented in the \texttt{fipp} \texttt{R} package \citep{greve2021package}.

\subsection{Repulsive prior distributions}

To reduce the number of estimated clusters, there are various approaches that can be taken. One approach is to modify the prior distribution for the number of components or the partition. Additionally, it is possible to define prior distributions for the component parameters in a way that favors well-separated components. When assuming independent and identically distributed parameters a priori, components can be randomly located in the parameter space, potentially leading to components that are very close to each other. On the other hand, by introducing a repulsive prior distribution, dependence is introduced a priori among the parameters of a mixture model, particularly among the location parameters. This results in the parameters no longer being conditionally independent.

To incorporate repulsion, methods such as those proposed by \cite{quinlan2018density} and \cite{xie2020bayesian} include a penalization term based on pairwise distances between the location parameters. Another approach, suggested by \cite{malsiner2017identifying}, is to use repulsive mixtures, which encourage components to merge into groups at one hierarchical level while separating groups at another level. The properties of these prior distributions are further studied by \cite{quinlan2021class}.
 
A repulsive distribution can be expressed as $\text{Rep}_{K}(\theta_{K}) = \frac{1}{Const_{K}} \left\{ \prod_{h=1}^K f_0(\theta_h)\right\} R_C(\theta_{K})$, where $f_0$ is a probability density function and $Const_{K}$ is a normalizing constant. The function $R_C(\theta_K)$ is defined as
$$
R_C(\theta_K) = \prod_{1 \leq r \leq s \leq K} [1 - C_0\{\rho(\theta_r, \theta_s)\}], 
$$
where the function $C_0: [0,\infty) \rightarrow (0,1]$ satisfies the following conditions: i) $C_0(0)=1$, ii) $C_0(x) \rightarrow 0$ as $x \rightarrow \infty$, and iii) for any $x, z \geq 0$, if $x < z$ then $C_0(x) > C_0(z)$. The function $C_0$ is associated with the potential $\phi$, given by
$$
\phi(\theta_r,\theta_s) = -\log \{1 - C_0(\rho(\theta_r, \theta_s))\}.
$$ 
By assuming a repulsive prior for the location parameters of a mixture model, an interaction structure among them is induced, which is defined by $\rho(\theta_r,\theta_s)$ through $C_0$. Various repulsive distributions have been proposed in the literature. For instance, \cite{ogata1985estimation} introduce a repulsive distribution associated with soft repulsion, while \cite{petralia2012repulsive} present a repulsive distribution with a potential, resulting in stronger repulsion. The choice of the repulsive distribution has an impact on the estimated number of clusters. Soft repulsion tends to eliminate singletons while maintaining good density estimation properties, whereas strong repulsion leads to a higher level of parsimony, with only a small number of estimated clusters.

The function $\rho(\cdot, \cdot)$ can be chosen as the Mahalanobis distance, and $C_0(a) = \exp\left(-\frac{1}{2}\frac{a^2}{\tau}\right)$, where $\tau$ is a parameter controlling the strength of repulsion. When $\tau \rightarrow 0$, the repulsion is weaker, while $\tau \rightarrow \infty$ leads to stronger repulsion. The parameter $\tau$ can be either estimated or fixed. Treating $\tau$ as an unknown parameter and assigning it a prior distribution significantly increases the computational burden and reduces the tractability of the posterior distribution. However, fixing $\tau$ may strongly influence the type of repulsion implied.

A repulsive prior distribution on the location parameters of a mixture model induces a prior distribution on the number of clusters, but its explicit expression is not available. \cite{quinlan2021class} prove that when the true cluster locations are separated by a minimum Euclidean distance that favors distinct clusters, and the prior assigns positive mass to arbitrarily small neighborhoods around the true density, the posterior rate of convergence relative to the $L_1$-metric is $\varepsilon_n = n^{-1/2} \log(n)^{1/2}$.

Alternatively, instead of repulsive priors, \cite{fuquene2019choosing} propose a ``non-local prior'' approach for selecting the number of components. A non-local prior distribution for the model with $k$ components assigns vanishing probability as the mixture with $k'$ components becomes equivalent to a mixture with $k$ components when $k=k'$ or $\theta_h = \theta_j$ for some $h \neq j$. This prior distribution only requires identifiability of the model, meaning that $g(y | \theta_k, K=k) = g(y | \theta_{k'}, K=k')$ only when $k=k'$ and $\theta_k = \theta_{i(k')}$ for some permutation $i(k')$ of the component labels in the model with $k'$ components. However, \cite{fuquene2019choosing} demonstrate through simulation studies and real datasets that this approach may be overly conservative, resulting in an underestimation of the number of components. Nevertheless, it appears to be more robust than other approaches to misspecifications of the component distributions \citep{cai2021finite}.

The suitability of repulsive mixtures relative to static or dynamic mixture models depends on the specific application. For example, \cite{quinlan2018density} apply both dependent Dirichlet processes (see Section \ref{sub:DDPcov}) and repulsive mixture models to a regression density problem to study the relationship between the duration and waiting time of eruptions from the Old Faithful geyser in Yellowstone National Park, Wyoming. They show that repulsive mixtures tend to be more parsimonious, making them suitable for well-separated clusters. However, in the context of multivariate data, a larger number of clusters might be needed to accurately represent partitions in higher dimensions, where repulsive mixture models may be too parsimonious.

\subsection{Infinite mixture models}
\label{sub:dpmixtures}

\subsubsection{Dirichlet processes.} 
An alternative to the model given in Equation \eqref{eq:finitemix} is to consider a model with an infinite number of components. One of the main tools used in this context is the Dirichlet process. Let $F$ be a random probability measure on $\Theta$. We write $F \sim DP(\alpha, F_0)$ to denote that $F$ is distributed according to a Dirichlet process with base measure $F_0$ and concentration parameter $\alpha$.

The success of DPs in Bayesian analysis is due to their conjugacy: if $F\sim DP(\alpha,F_0)$, then the posterior distribution is also a Dirichlet process. 
It can be proven \citep{Ferguson1973} that the support of the random variable $F$ is almost surely the family of discrete distributions. Although this limits its density estimation properties, this feature introduces a clustering ability in the DP, as demonstrated by the P\'olya urn scheme of \cite{blackwell1973ferguson}.
Consider a sequence of independent and identically distributed (i.i.d.) variables $\theta_1, \theta_2, \ldots \sim F$ where $F \sim DP(\alpha, F_0)$. Since $\theta_{n+1} | F, \theta_1, \ldots, \theta_n \sim F$, 
the predictive distribution of $\theta_{n+1}$ given $\theta_1, \ldots, \theta_n$ is
\begin{equation*}
    \theta_{n+1} | \theta_1, \ldots, \theta_n \sim \frac{1}{\alpha+n} \left( \alpha F_0 + \sum_{i=1}^n \delta_{\theta_i}\right).
\end{equation*}
This predictive distribution is a mixture of the base distribution $F_0$ and the empirical distribution of the atoms already drawn; here, $\delta_x$ represents the Dirac mass at $x$. In other words, it contains point masses at $\theta_1, \ldots, \theta_n$, which implies that there are ties in a sequence of $n$ sequentially drawn atoms. 
For further information, refer to \cite{aldous1985exchangeability} and \cite{pitman2002}, which discuss a similar construction based on random partitions.

\subsubsection{Dirichlet processes mixture models.} 
Mixing a Dirichlet Process (DP) with respect to kernels results in a countable mixture of distributions \citep{antoniak1974}. Consequently, it is possible to model a set of random variables $Y_1, \ldots, Y_n$ (which can be possibly absolutely continuous) using the atoms of the Dirichlet process as latent parameters $\{\theta_1, \ldots, \theta_n\}$. The model can be represented as in Equation \eqref{eq:partmod1}, with the discrete random probability measure $F$ defined by a $DP(\alpha, F_0)$ prior process.
 
Since $F$ is almost surely discrete, this model can be rewritten as
\begin{equation*}
    Y_i \sim \sum_{h=1}^{\infty} \pi_h f(y_i|\theta_h^*) \qquad i=1, \ldots, n,
\end{equation*}
where $\theta^*_1,\theta^*_2, \ldots$ are independent draws from the base distribution $F_0$. Thus, the clustering properties of the DP naturally extend to the case of DP mixture models.

The DP weights $\pi_h$ can be constructed using a stick-breaking process \citep{sethuraman:stick}: $\pi_1 = V_1$ and $\pi_h = V_h \prod_{\ell<h} (1-V_{\ell})$, where $V_1, V_2, \ldots$ are independent beta random variables $V_h \sim Be(a_h,b_h)$. Alternatively, the probabilities $\pi_h$ can be drawn from any distribution on the simplex, as described in \cite{ongaro2004discrete}.
The DP prior is obtained when $V_h \sim Beta(1,\alpha)$ for all $h$. The Pitman-Yor (PY) process is obtained when $V_h \sim Be(1-\gamma_a, \gamma_b+\gamma_a h)$ for $\gamma_a \in [0,1)$ and $\gamma_b > -\gamma_a$ for all $h$. When $\gamma_a > 0$, the expected value $\mathbb{E}[V_h]$ is a decreasing function of $h$. Consequently, sets with larger mass $\pi_h$ are typically associated with smaller indexes $h$. This characteristic explains why the resulting partition follows heavy-tailed power-law distributions \citep{goldwater2006interpolating}. In this regard, the PY process is more effective in estimating the number of rare or small clusters compared to a DP. When $\gamma_a = 0$, the DP is recovered with a concentration parameter of $\gamma_b$.

The relationship between finite and infinite mixture models has also been recently studied by \cite{argiento2019infinity}, who provide a unifying framework based on a construction using normalization of point processes. \cite{argiento2019infinity} also characterize the induced partition of the proposed process.

An interesting extension of the DP is proposed by \cite{wade2011enriched}. The main idea is that each atom is a random vector that can be partitioned into two groups, with the sample space represented as the product of two finite spaces. The joint random distribution is then modeled in terms of marginal and conditional distributions, with independent DP priors assigned to each term. This resulting process is more flexible than a DP, with additional parameters determining the variability.

\subsubsection{EPPF and PPF of Dirichlet and Pitman-Yor processes.} 
Even though the number of components in the DP (or any generalization of it) is infinite, only a random finite subset of components has a probability $\pi_h$ greater than zero, denoted as $K_{+} = \{h: \pi_h > 0\}$. Hence, it is possible to infer the number of underlying clusters, with the probability of assigning an observation to the $h$-th cluster given by $\pi_h$.

The explicit expression for the EPPF of the DP is available. If $F \sim DP(\alpha,F_0)$, then
\begin{equation*}
P(\rho_n = (S_1, \ldots, S_{K_+})) = p(|S_1|, \ldots, |S_{K_+}|) = \frac{\alpha^{K_+} \prod_{h=1}^{K_+} \left( n_h-1\right)!}{\prod_{i=1}^n (\alpha+i-1)!},
\end{equation*}
which is known as the Ewens distribution. This equation can be generalized to the case of the Pitman-Yor process (PY) as follows:
$$
P(\rho_n = (S_1, \ldots, S_{K_+})) = p(|S_1|, \ldots, |S_{K_+}|) = \frac{\left( \prod_{h=1}^{K_+ -1} (\alpha + h\gamma_a)\right) \left( \prod_{h=1}^{K_+} (1-\gamma_a)^{(n_h-1)}\right)}{(1+\alpha)^{n-1}},
$$
which is known as the Ewens-Pitman distribution. Other Gibbs-type priors also have explicit EPPFs \citep{gnedin2006exchangeable, ho2007gibbs, lijoi2008bayesian, gnedin2010species, cerquetti2013marginals, de2013gibbs}.

There is a second family of Pitman-Yor (PY) processes, where $b < 0$ and $a = K_+|b|$ with $K_+ \in \mathbb{N}$ and known \citep{gnedin2010species,de2013gibbs}. The EPPF of this representation of the PY process is given by:
$$
p(\rho_n = (S_1, \ldots, S_{K_+})) = p(|S_1|, \ldots, |S_{K_+}|) = \frac{\Gamma(a)}{\Gamma(n+a)} \prod_{h=1}^{K_+} (a + b(h-1)) \frac{\Gamma(n_h - b)}{\Gamma(1-b)}. 
$$
A static finite mixture with $K$ components and a symmetric Dirichlet prior on the weights, with all parameters equal to $\gamma$ (fixed and known), is obtained by mixing a $PY(-\gamma, K\gamma)$ process prior over the concentration parameter $\alpha = K \gamma$, and fixing $b = -\gamma$ \citep{gnedin2006exchangeable}. On the other hand, the dynamic finite mixture model is derived by mixing a $PY\left(-\frac{\alpha}{K}, \alpha\right)$ prior over $b = -\frac{\alpha}{K}$, while the concentration parameter is fixed at $a = \alpha$. 

Also, the PPF of a DP is available in closed form, which is given by the P\'olya urn representation \citep{blackwell1973ferguson}:
\begin{equation}
\label{eq:PPFEwens}
p_j(n_1, \ldots, n_{K_+}) = \left( \frac{n_j}{n+\alpha} \right) \mathbb{I}\{1 \leq j \leq K_+\}  + \left( \frac{\alpha}{n+\alpha}\right) \mathbb{I}\{j =  K_++1\}.
\end{equation}
This expression can be generalized to the PY process $PY(\gamma_a,\gamma_b,F_0)$ as follows:
\begin{equation}
\label{eq:PPFEP}
p_j(n_1, \ldots, n_{K_+}) = \left( \frac{n_j-\gamma_a}{n+\gamma_b} \right) \mathbb{I}\{1 \leq j \leq K_+\}  + \left( \frac{\gamma_b + K_+ \gamma_a}{n+\gamma_b}\right) \mathbb{I}\{j =  K_++1\}.
\end{equation}

The conditional EPPF for a DP mixture model, induced for a given number of clusters $K_+ = k$, is given by:
$$
p_{DP}(\rho_n = (S_1, \ldots, S_{k}) | K_+ = k) = p(|S_1|, \ldots, |S_{K_+}| | K_+ = k) =  \frac{1}{Const_{\infty}} \prod_{h=1}^k \frac{1}{n_h},
$$
where $Const_{\infty}$ is the normalizing constant taken as the summation with respect to all the labeled cluster sizes for which the sum is equal to $n$. This conditional EPPF favors unbalanced partitions with some small values of $n_h$ due to the inverse dependence on $n_h$, for $h=1, \ldots, k$ \citep{antoniak1974, miller2018mixture}. More precisely, this conditional EPPF has the form of a discrete Dirichlet distribution for a finite mixture model when $K_+ = k$, while it is an improper Dirichlet distribution (with all parameters equal to zero) in the case of an infinite mixture model. This means that as $n \rightarrow \infty$, the distribution of $(S_1, \ldots, S_k) | K_+ = k$ concentrates all its mass at the corners of the simplex in the case of a DP mixture model.

\subsubsection{Inconsistency of the posterior distribution on the number of clusters.} 
For DP mixture models, \cite{antoniak1974} also provides the induced prior distribution on $K_+$ as $p_{K_+}(k_+) = \frac{\Gamma(\alpha)}{\Gamma(n + \alpha)} \mathcal{S}_{n, k_+}$ where $\mathcal{S}_{n, k_+}$ is the Stirling number of the first kind. The number of clusters grows as $K_+ \sim \alpha \log (n)$. This reveals another perspective showing that the DP mixture model favors the estimation of many small clusters \citep{antoniak1974,argiento2009comparison, onogi2011characterization} when the concentration parameter is fixed.

These results have led \cite{miller2014inconsistency} to analytically prove that the posterior distribution on the number of clusters does not concentrate on any finite value as the sample size $n$ increases: 
$$
\lim \sup_{n \rightarrow \infty} P(K_+ = k | y_1, \ldots y_n) < 1
$$
with probability one. 
This holds for a large class of models, including DP and PY processes with components from a broad range of distribution families. 
Suppose $\rho_n^k$ is a partition with $k$ components, and define $\rho_n^{k'}$ as a partition with $k' = k+1$ components. Then $\rho^{k'}_n$ can be generated by splitting one element in $\rho_k$ to be in its own cluster (a singleton). The reason behind these results lies in the fact that the likelihood for a model with $k$ clusters, $f(y_1, \ldots, y_n | (S_1, \ldots, S_k))$, is of the same order as the likelihood for a model with $k' = k+1$ clusters (where one observation is removed from an existing cluster to create a singleton), $f(y_1, \ldots, y_n | (S_1, \ldots, S_{k'}))$. However, the induced prior distribution on the partition provided by a PY process tends to favor models with additional clusters. Therefore, a PY mixture model tends to give preference to models with small clusters. \cite{alamichel2022bayesian} extend these inconsistency results to other Bayesian nonparametric priors, such as Gibbs-type processes and their finite-dimensional representations, including the Dirichlet multinomial process and the Pitman–Yor and normalized generalized gamma multinomial processes. One can see this problem as a misspecified model where a model with a finite number of clusters is represented by a model with an infinite number of components.

An alternative solution is to use a post-processing algorithm. \cite{guha2021posterior} propose a post-processing Merge-Truncate-Merge (MTM) algorithm that enables consistent inference of the number of clusters in mixture models, specifically focusing on DP mixtures. \cite{alamichel2022bayesian} extend this MTM approach to PY process mixtures and overfitted mixtures. When the Wasserstein convergence rate of the mixing measure is known, the MTM algorithm provides a consistent estimate of the number of components. However, \cite{alamichel2022bayesian} suggest that challenges still exist when estimating a finite number of components using Gibbs-type process mixture models, as these models inherently assume an infinite number of components or a number of clusters that grows with the sample size.

On the other hand, \cite{ascolani2023clustering} examine DP mixtures and determine the consistency of cluster number estimation when data originate from a finite mixture and a prior is placed on the DP's concentration parameter, in contrast to \cite{miller2014inconsistency}, where the concentration parameter is fixed. While the assumptions about the prior distribution for the concentration parameter in \cite{ascolani2023clustering} are mild and are satisfied by commonly used priors (such as uniform or gamma distributions), there is a need for the kernel representing the component distribution to be perfectly specified. Otherwise, phenomena similar to those observed by \cite{cai2021finite} for finite mixture models are likely to occur. Additionally, a condition of separability of the atoms in the true data-generating process is required, meaning that cluster locations must be sufficiently distinct to be recognized. Furthermore, these results have not yet been extended to dependent DP (DDP), which are known to estimate a larger number of clusters compared to the standard DP \citep{grazian2023spatio}.

\subsubsection{Other parameters.} 
Finally, the parameters of the base distribution $F_0$ and the concentration parameter $\alpha$ can also be considered as random variables with their own prior distribution. In particular, $\alpha$ plays a crucial role in the distribution induced on the number of clusters and partitions \citep{ascolani2023clustering}. Several works suggest using a Gamma distribution, $\alpha \sim \text{Ga}(a,b)$ \citep{Escobar1995,jara2007dirichlet}. The standard ``non-informative'' choice of setting $a$ and $b$ to be close to zero results in a highly informative prior for the number of clusters, with concentration around one and infinity \citep{dorazio2009selecting}. An alternative proposal by \cite{fruhwirth2021generalized} is to use the $F$-distribution, $\alpha \sim \mathcal{F}(\nu_l, \nu_r)$, where the parameters control different characteristics of the prior distribution. A small $\nu_r$ gives fat tails, while a small $\nu_l$ tends to favor models with a small number of clusters.

\subsubsection{Computational aspects.} 
Computationally, \cite{ishwaran2001gibbs} propose computational methods for the most general case of $V_h \sim \text{Beta}(a_h, b_h)$. Given the stick-breaking representation of the DP, there are two main ways to perform posterior inference through Gibbs sampling. The first one is associated with the P\'olya urn Gibbs sampler \citep{Escobar1994,maceachern1994estimating,Escobar1995,maceachern1998computational}. The second one is the blocked Gibbs sampler proposed by \cite{ishwaran2001gibbs}. lternatively, a popular algorithm used for DP mixture models is the slice sampling method proposed by \cite{kalli2011slice}.

\subsection{Dirichlet process mixtures in presence of covariates}
\label{sub:DDPcov}

DPs are based on the assumption that data are infinitely exchangeable, meaning that the ordering of data items does not matter. However, this assumption can be unrealistic, and many works have attempted to model more structured data. In particular, it is important to define a model in which the distribution $F_x$ changes smoothly with respect to $x \in \mathcal{X}$, such that $F_{x_1} \rightarrow F_{x_2}$ as $x_1 \rightarrow x_2$. The dependent Dirichlet process (DDP) \citep{maceachern2000dependent} is a generalization of the DP that creates a distribution on the set of countable mixture distributions. DDP introduces dependence among collections of distributions, where the dependence is driven by a covariate $x$. This is achieved by allowing the atoms $\theta_h^*$ to be replaced by a process $\theta_{\mathcal{X}}$, which provides the atom for each value of the covariate. Similarly, the random variable $V_h$ in the stick-breaking construction can be replaced by a process $V_{\mathcal{X}}$, which determines the mass assigned to $\theta_{\mathcal{X}}$ at each level of the covariate. For recent reviews on this topic, the reader is referred to \cite{quintana2020dependent} and \cite{wade2023bayesian}. While highly investigated for density estimation, these processes are much less studied for clustering (while often used in practice for that purpose) and analytical expressions of the EPPF may not be available. 

\subsubsection{Single-$\pi$ dependent Dirichlet processes.} An important class of DDPs is the single-$\pi$ DDP, which offers a significant simplification in terms of computation. The key idea behind the single-$\pi$ DDP is that the mass $\pi_h(x)$ does not vary with $x$. As a consequence of this restriction, the model can be regarded as a countable mixture of stochastic processes, with mixing weights that align with those of a single DP model. The single-$\pi$ DDP is useful for smoothing the prediction distribution across the covariate space. However, it is not suitable for clustering tasks because the DP probabilities are not dependent on the covariate. Single-$\pi$ DDP have been successfully applied to regression problems in the ANOVA DDP \citep{de2004anova}, such as in survival analysis, and \cite{de2007semiparametric} use it for longitudinal analysis.

\subsubsection{Single-$\theta$ dependent Dirichlet processes.} Several authors have proposed an extension of the stick-breaking construction introduced by \cite{sethuraman:stick} that allows the probabilities $\pi_h$ to vary with the covariate. For example, \cite{reich2007multivariate}, \cite{dunson2008kernel}, \cite{warren2012bayesian}, and \cite{grazian2023spatio} have explored this idea. In this extended framework, the random variable depending on covariates follows a model given by:
\begin{equation}
	Y_i|x_i \sim \sum_{h=1}^\infty \pi_h(x_i)\delta_{\theta^*_h} \qquad i=1, \ldots, n.
	\label{eq:single-atom}
\end{equation}
The weights $\pi_h(x_i)$ are constructed using a stick-breaking process, where $\pi_1(x_i) = V_1(x_i)$ and $\pi_h(x_i) = V_h(x_i)\prod_{j=1}^{h-1} (1-V_j(x_i))$ for $h>1$. However, in this extended construction, the variables $V_h(x_i)$ are allowed to vary according to a kernel function that smooths over the covariate space. Specifically, $V_h(x_i) = w_h(x_i)V_h$, where $V_h \sim \text{Beta}(a_h,b_h)$ and $w_h$ is a kernel function that is constrained within the interval $[0,1]$. This formulation incorporates dependence in the allocation probabilities, defining clusters characterized by kernel functions.
 
The model described in Equation \eqref{eq:single-atom} represents a case of the so-called ``single-atom'' DDP, where the atoms $\theta^*_h$ are independent with a marginal distribution $F_0$. \cite{fuentes2013multivariate} and \cite{grazian2023spatio} further extend this model by incorporating dependence among the atoms $\theta^*(x)$ of the DP using a Gaussian process as the base distribution.

The dependence imposed on the probabilities $\pi_1(x), \pi_2(x), \ldots$ can also be described through a model. \cite{chung2009nonparametric} propose a construction that relies on a probit representation of the variables used to construct the clustering probabilities, instead of using beta random variables. On the other hand, \cite{papageorgiou2015bayesian} propose to directly model the mixture weights through a probit model. \cite{ren2011logistic} propose incorporating dependence on the weights of the mixture components through a logistic regression, via the use of a kernel depending on the distance between covariates level. Another possible construction uses geometric weights  \citep{fuentes2009nonparametric}. \cite{griffin2006order} define the mixing weights as transformations of i.i.d. random variables. However, they introduce dependence by inducing an ordering of the i.i.d. random variables at each covariate level, so that distributions at similar covariate levels are associated with similar orderings.

\cite{sudderth2008shared} extend the PY process to incorporate information provided by a covariate through a latent variable with a thresholded Gaussian distribution. The priors on the stick-breaking proportions $V_h \sim \text{Beta}(1-\gamma_a,\gamma_b + h\gamma_a)$ are then transformed into corresponding random thresholds.
When $\gamma_a = 0$, the model of \cite{duan2007generalized} is formally recovered. \cite{rodriguez2010latent} propose a latent stick-breaking process in which observations at different spatial locations are dependent but share a common marginal distribution.

Single-$\theta$ DDP have become popular for introducing spatial or temporal dependence—see, for example, Section \ref{subsub:spatial} — thanks to their ease of computation. Spatio-temporal dependence can be flexibly introduced via the use of a kernel, rather than defining a Gaussian process on the atoms of the DDP.

\subsubsection{Difficulties in introducing dependence on covariates.} While stick-breaking methods are appealing from a computational perspective, they encounter a natural difficulty. The stick-breaking construction of clustering probabilities involves transforming the variables on which the dependence is defined. This transformation modifies the structure of dependence, thereby making it challenging to control the dependence among the clustering probabilities. For a discussion of this problem in the context of finite mixture models, we refer the reader to \cite{mastrantonio2019new}. The vector of probabilities in stick-breaking methods is compositional, which complicates the interpretability of the dependence structure. Since the elements of a compositional vector are defined on the simplex, the covariance between each element $h$ and the sum of all elements in any finite $K$-dimensional sequence of the DP is given by $\mbox{Cov} ( \pi_{h} , \pi_{1} + \cdots + \pi_{h} + \cdots + \pi_{K}) = 0$. This is because $\pi_{1} + \cdots + \pi_{h} + \cdots + \pi_{K} = 1$. Consequently, we have:
\begin{equation*}
-\mbox{Var}(\pi_{h})=\sum_{\substack{ \ell=1  \\ h \neq \ell}}^{K}\mbox{Cov}(\pi_{\ell},\pi_{h}).
\end{equation*}
In other words, at least one element on the right side of the equation must be negative, and correlations are not allowed to vary freely in the range $(-1,1)$. Therefore, the sum-to-one constraint for any sub-sequence of the DP induces negative correlations among the probabilities \citep{Aitchison1986}. Recently, \cite{ascolani2023nonparametric} have derived a class of dependent nonparametric priors that can induce correlations of any sign, not necessarily negative, between the random probabilities, as well as across samples.

In more detail, let $(\pi_1(x), \pi_2(x), \pi_3(x))^T$ be a subsequence of a DP. By definition, such a vector follows a Dirichlet distribution. Consider two values of the covariate, $x_1$ and $x_2$, and suppose the beta variables of the stick-breaking construction are described by the following matrix:
\begin{equation*}
\left(
\begin{array}{cc}
 v_1(x_1) & v_2(x_1) \\
 v_1(x_2) & v_2(x_2) 
\end{array}
\right).
\end{equation*}
Let $p_1$ and $P_1$ denote the probability density function and the cumulative distribution function of $V_1(x)$, respectively. Similarly, let $p_2$ and $P_2$ denote the probability density function and the cumulative distribution function of $V_2(x)$. The joint cumulative distribution function of $(V_1(x_1),V_1(x_2))$ can be defined using a copula representation $C_1$ with density $c_1$. Similarly, define $C_2$ and $c_2$ for the bivariate random variable $(V_2(x_1), V_2(x_2))$. Then, the joint density of the four variables can be written as:
\begin{align*}
 p_{\mathbf{V}(\mathbf{x})}
 &\left(
\begin{array}{cc}
 v_1(x_1) & v_2(x_1) \\
 v_1(x_2) & v_2(x_2) 
\end{array}
 \right)= 
 c_1
 \left(
 \begin{array}{c}
 P_1( v_1(x_1))\\
 P_1( v_1(x_2))\\
  \end{array}
 \right)
 p_1(v_1(x_1))  \; p_1(v_1(x_2)) \cdot \\
 &\cdot c_2
 \left(
 \begin{array}{c}
 P_2( v_2(x_1))\\
 P_2( v_2(x_2))\\
 \end{array}
 \right)
 p_2(v_2(x_1))  \; p_2(v_2(x_2)).
\end{align*}
where $\mathbf{V}(\mathbf{x})$ stands for a $2\times 2$ matrix with elements $[V_j(x_i)]_{ji}$ for $j=1,2$, $i=1,2$.
 
The vector $[V_1(x), V_2(x), \ldots]$ represents the vector on which the dependence is constructed. However, the focus is on the corresponding vector $[\pi_1(x), \pi(x), \ldots]$, which is associated with $[V_1(x), V_2(x), \ldots]$ through a one-to-one map from the space $\mathcal{V}^2$ to $[0,1]^3$. This mapping is achieved using a matrix of transformations $B$:
\begin{align*}
 B = \left[\begin{array}{cc}
 \pi_1(x_1) = v_1(x_1) &\pi_1(x_2) = v_1(x_2) \\
 \pi_2(x_1) = v_2(x_1)(1-v_1(x_1))& \pi_2(x_2) = v_2(x_2)(1-v_1(x_2)) \\
  \pi_3(x_1) = 1-\pi_1(x_1) - \pi_2(x_1)& \pi_3(x_2) = 1-\pi_1(x_2) - \pi_2(x_2)
 \end{array}
  \right].
  \end{align*}
The two-dimensional space of $(V_1, V_2)$ is expanded to the three-dimensional space of $(\pi_1, \pi_2, \pi_3)$. By employing a change of variables, it becomes possible to derive the joint density of the probability vectors
\begin{small}
 \begin{align*}
p\left(
\begin{array}{ccc}
\pi_1(x_1) & \pi_2(x_1) & \pi_3(x_1)\\
 \pi_1(x_2) & \pi_2(x_2) & \pi_3(x_2)
 \end{array}
\right) = 
c^*
\left(
\begin{array}{ccc}
P_1( \pi_1(x_1)) & P_2\left(\frac{\pi_2(x_1)}{1-\pi_1(x_1)}\right) & P_g(g^{-1}(\pi_3(x_1)))\\
P_1( \pi_1(x_2)) & P_2\left(\frac{\pi_2(x_2)}{1-\pi_1(x_2)}\right) & P_g(g^{-1}(\pi_3(x_2)))\\
\end{array}
\right) \cdot \\
\cdot p_1(\pi_1(x_1))  p_1(\pi_1(x_2)) 
p_2\left(\frac{\pi_2(x_1)}{1-\pi_1(x_1)} \right)   p_2\left(\frac{\pi_2(x_1)}{1-\pi_1(x_1)} \right) \cdot p_g(g^{-1}(\pi_3(x_1))) h_g(g^{-1}(\pi_3(x_2))) \cdot | \mathbb{J}(B^{-1})|.
\end{align*}
\end{small}
Here, $c^*$ represents the copula of the augmented three-dimensional variable, and $|\mathbb{J}(B^{-1})|$ denotes the determinant of the Jacobian matrix of the inverse one-to-one map. The determinant of the Jacobian matrix depends on functions of $(\pi_1(x_1), \pi_2(x_1), \pi_1(x_2), \pi_2(x_2))$, given the deterministic definition of $(\pi_3(x_1), \pi_3(x_2))$. As a result, the structure of the dependence, described by the copula density $c^*(\cdot)$, is altered.

\begin{figure}[h]
    \centering
    \includegraphics[height=8cm,width=12cm]{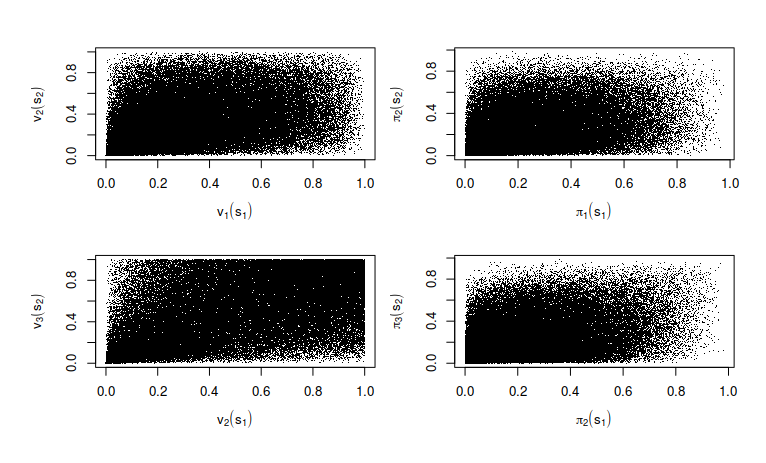}
    \caption{Experiment to show the change in the dependence structure implied by the stick-breaking construction: on the left side the beta random variables are shown, on the right side the probabilities obtained through the stick-breaking transformation of the beta variables are shown.}
    \label{fig:cop}
\end{figure}

To illustrate this characteristic, Figure \ref{fig:cop} presents the results of a simulation involving $V_h(x)$ drawn from beta distributions, with dependence expressed through a Clayton copula parameterized at one. These variables have been transformed into compositional vectors using the stick-breaking construction. The scatterplots resulting from $10^5$ simulations are displayed in Figure \ref{fig:cop}. The left side depicts the variables $V_1(x)$ and $V_2(x)$, while the right side illustrates the variables $\pi_1(x)$ and $\pi_2(x)$. As $\pi_3(x)$ is deterministically derived, it is not shown. It is apparent that the dependence structure originally present in the beta random variables is not preserved in the probabilities.

\subsubsection{Distance dependent Chinese restaurant process.} Alternatively to DDP, \cite{blei2011distance} introduce the distance dependent Chinese restaurant process (ddCRP), which directly models the probability of assigning observations to available clusters. The underlying assumption of the ddCRP is that data points that are close to each other, based on some form of distance, are more likely to be clustered together. Consequently, the assignment is based on the distances between observations, connecting each observation with others rather than with the atoms of the DP.

Let $C_i$ denote the assignment for the $i$-th observation, and $d_{ij}$ be a distance measure between observation $i$ and observation $j$. $D$ represents the set of all distance measurements between observations, and $m$ is a decay function. The probability of assigning observation $i$ to cluster $j$ given the distances ${d_{ij}}$, the decay function $m$, and the concentration parameter $\alpha$, is proportional to:
$$
p(C_i = j | \{d_{ij}\}, m, \alpha) \propto \begin{cases}
m(d_{ij}) 	& 	j \neq i \\
\alpha		&	j = i.	
\end{cases}
$$

The assignment of an observation depends solely on the distance $d_{ij}$, which can represent various measures such as time difference, Euclidean distance for spatial points, and so on. The choice of $m(\cdot)$ determines the behavior of the process and should possess several key properties, including being non-increasing, non-negative, having finite values, and satisfying $m(\infty) = 0$.

Additionally, \cite{ghosh2011spatial} define a hierarchical version of the ddCRP that clusters groups of observations. This hierarchical version allows for sharing of cluster components across groups, and within-group clustering depends on distances among locations.

\subsubsection{Dirichlet process mixtures for spatial data.} \label{subsub:spatial}

DP mixtures have also been widely applied to spatial data. The formal definition presented in the previous sections remains valid, where the spatial location serves as the covariate. For instance, in the work of \cite{gelfand2005bayesian}, the atoms $\theta_h$ are allowed to depend on the location $s$, denoted as $\theta_h = \theta_h(s)$. Assume that replications are available at the $J$ spatial location, therefore $Y_i = (Y_i(s_1), \ldots, Y_i(s_J))$, for $i=1,\ldots, n$. A DP is placed on $F^{(J)}$, $F^{(J)} \sim DP(\alpha, G_0^{(J)})$, where $G_0^{(J)}$ is a mean-zero $J$-multivariate normal distribution with covariance matrix that can be chosen to have stationary and possibly isotropic correlation function. Then the atoms $(\theta_{i}(s_1), \ldots, \theta_i(s_J))$, for $i=1,\ldots,n$ are i.i.d. realisation from $F^{(J)}$.

The distribution of the $i$-th observation can then be represented as  
\begin{equation*}
(Y_i(s_1), \ldots, Y_i(s_J)) \sim \sum_{h=1}^{\infty} \pi_h \delta_{\theta^*_h(s1),\ldots, \theta^*_h(s_J)}.
\end{equation*}
where $(\theta_h^*(s_1), \ldots, \theta_h^*(s_J))$ are realisations from the Gaussian process $G_0^{(J)}$.

Both \cite{duan2007generalized} and \cite{gelfand2007bayesian} extend the spatial Dirichlet Process (SDP) by considering varying mixture weights, which allow observations to depend on different surfaces at different locations. Here, the weights can be defined to allow for site-specific selection of surfaces, enabling similar weights to be assigned to sites that are close together. It is worth noting that due to the lack of techniques for deriving a posterior distribution for the weights, this process is not suitable for clustering purposes, which is a task studied in the extension proposed by \cite{petrone2009hybrid}. 

One of the main limitations of the spatial DP is that it requires replications at each spatial location. Consequently, these models are not suitable when the data represent only one surface. To address this limitation, \cite{reich2007multivariate} propose a spatial extension of the stick-breaking construction that allows the probabilities $\pi_h$ to vary spatially without relying on the presence of replications. This model is a case of single-$\pi$ process, where probabilities are defined through a kernel function smoothing over space. \cite{grazian2023spatio} proposes a spatio-temporal extension of this model, also comparing a single-$\theta$ model with a model where both weights and atoms are allowed to depend on space and time. While the conceptual introduction of dependence on both the atoms and the weights of the process may seem straightforward, the computational time required drastically increases.

Another computational limitation of the kernel stick-breaking prior is related to the dimensionality of the data. To make the algorithm feasible, these approaches usually select a limited area around each point location. In order to perform dimension reduction, an approach developed by \cite{reich2012variable} based on Bayesian variable selection can be employed, where only informative spatial locations are included in the definition of the kernels.

Although the method implemented by \cite{reich2007multivariate} is designed for point-referenced data, generalizations to account for areal data have also been proposed, such as the areally-weighted stick-breaking process and the areally-referenced DP of \cite{li2010nonparametric}. Similar methodologies can be applied to model spatial-varying regression coefficients \citep{cai2013bayesian}. 

An interesting property often sought in modeling spatial data is that spatial clusters should consist of adjacent areal units. \cite{wehrhahn2020bayesian} propose a modified EPPF associated with the Ewens distribution, called the restricted Chinese restaurant process, which enforces clusters of adjacent units. The main idea is to set the prior distribution to assign zero probability to configurations where clusters consist of non-adjacent units. This is achieved by introducing a latent graph where adjacency is defined by edges in the subgraphs. A typical application of this approach is in epidemiology, where data come from health districts, and the goal is to cluster these districts under the hypothesis that communicable diseases spread in proximity.

\section{Prior distributions on the partitions}
\label{sec:ppm}

Using an induced model on the partitions may have some drawbacks: spatial correlation can sometimes be counter-intuitive \citep{wall2004close}, and local features may not be adequately captured. Alternatively, it is possible to directly define a prior distribution on the partition by re-expressing a Gibbs-type prior model as follows:
\begin{align*}
Y_i | \theta^*_{C_i}, \rho_n &\stackrel{ind}{\sim} g(y_i | \theta^*_{C_i}) \qquad i=1, \ldots, n \\
\theta^*_{1}, \ldots, \theta^*_{K} &\stackrel{i.i.d.}{\sim} F_0 \\
C_i = h & \Leftrightarrow i \in S_h  \qquad i=1, \ldots, n \text{ and } h=1, \ldots, K\\
\rho_n &\sim p(\rho_n).
\end{align*}
In this model, the prior distribution $p(\rho_n)$ is defined directly on the partition. This means that the experimenter can directly control how prior information is transferred to a prior distribution. However, this construction does not guarantee sample size consistency and may not result in a valid EPPF. It is indeed possible to define a prior distribution on the partition that recovers the same EPPF as a DP, ensuring that the analyses correspond.

\subsection{Product partition models}

A popular method for defining a prior distribution directly on partitions is provided by product partition models \citep{hartigan1990partition,barry1992product}. In this approach, the prior distribution is given by:
\begin{equation*}
p(\rho_n = (S_1, \ldots, S_{K})) = c_0 \prod_{h=1}^{K} c(S_h). 
\end{equation*}
Here, $c(S_h) \geq 0$ is a non-negative function called the cohesion of $S_h$, which represents a measure of the strength of the prior assumption that elements in $S_h$ should be clustered together. The constant $c_0$ is a normalizing constant that sums over all possible partitions.

The prior distributions for the parameters of the model can be defined as:
$$
\pi(\theta^*_1, \ldots, \theta^*_{K}) = c_0 \sum_{(S_1, \ldots, S_{K}) \in \mathcal{P}_n} \prod_{h=1}^{K} c(S_h) \pi_h(\theta_h).
$$
If the subjects are exchangeable, product partition models are conjugate, and the posterior distribution $p(\rho_n | y_1, \ldots, y_n)$ is still a product model
$$
p(\rho_n = (S_1, \ldots, S_{K}) | y_1, \ldots, y_n) = c'_0 \prod_{h=1}^{K} c(S_h) f(y^*_h | \theta^*_{S_h})
$$
where $y^*_h = \{y_i, i \in S_h\}$. Additionally, product partition models have the property of sample size consistency, i.e. the prior distribution of the partition of the first $(n-1)$ indices can be obtained by integrating the joint distribution with respect to the $n$-th index.

\cite{quintana2003bayesian} demonstrate the connection between product partition models and DP mixture models. The P\'olya urn representation of the DP implies that the induced prior distribution for the partition $\rho_n$ is given by:
$$
p(\rho_n) = \frac{\alpha^K \prod_{h=1}^K \Gamma(|S_h|)}{\prod_{i=1}^n (\alpha + i - 1)} \propto \prod_{h=1}^K \alpha \Gamma(|S_h|)
$$
where $\alpha$ is the concentration parameter of the DP. This representation is proportional to the product over partition components, with $c(S) = \alpha \Gamma(|S|)$. Thus, the DP can be viewed as a product partition model, and as a result, product partition models yield a valid EPPF.

\subsection{Product partition models in presence of covariates}

\subsubsection{Using a similarity function.} 
The cohesion function $c(S_h)$ can be modified to include an additional regression function. Let $w(x_h^*)$ denote a nonnegative similarity function that measures the homogeneity of $x_i$ within cluster $S_h$, where $x_h^*=\{x_i, i \in S_h\}$. The modified expression for the prior distribution of the partition $\rho_n$ conditioned on the covariates $x_1, \ldots, x_n$ is:
$$
p(\rho_n | x_1, \ldots, x_n) = c_{0x} \prod_{h=1}^{K} c(S_h) \cdot w(x_h^*),
$$
where $c_{0x}$ is the normalization constant. The similarity function $w(\cdot)$ introduces a penalty for cluster size, such that $ \lim_{n_h\to\infty} w(x_h^*) = 0 $. To facilitate calculations, $w(\cdot)$ can be defined by marginalizing over an auxiliary model $q(\cdot)$:
$$
w(x_h^*) = \int \prod_{i \in S_h} q(x_i | \xi_h) q(\xi_h) d\xi_h,
$$
where $\xi_h$ represents a set of parameters of the auxiliary model. It is important to note that this representation does not necessarily imply that the covariates are random, but it is a convenient way to introduce correlations among similar values of the covariates. Under the assumptions of i) symmetry with respect to permutations and ii) scaling across sample size, which means the similarity of any cluster is the average of the augmented clusters, i.e., $w(x^*) = \int w(x^*, x) dx$, it can be proven that $w(x^*)$ is necessarily proportional to the marginal distribution of $x^*$ under a hierarchical auxiliary model. This is a direct application of De Finetti's representation theorem. In practical applications, the auxiliary model $q(\cdot)$ can be chosen in a conjugate form to evaluate the integral analytically.

In the case where $p$ covariates are available, the similarity function $w(\cdot)$ can be easily extended to include all covariates as $w(x^*_{h1}, \ldots, x^*_{hp}) = \prod_{\ell=1}^p q_{\ell}(x^*_{h\ell})$, where $x_{h\ell}^* = \{x_{i\ell}: i\in S_h\}$. This allows for incorporating multiple covariates into the similarity measure. The random partition model maintains coherence across different sample sizes when observations are independent across clusters and exchangeable within clusters.

\subsubsection{Using a covariate-dependent cohesion function.} 
\cite{park2010bayesian} generalize the previous class of product partition models by introducing a new definition:
$$
p(\rho_n = (S_1, \ldots, S_{K}) | x^*_1, \ldots, x^*_K) \propto \prod_{h=1}^{K} c(S_h, x^*_h).
$$
The posterior distribution of $(S_1, \ldots, S_K)$ remains a product partition model with an updated cohesion function. Similar to \cite{muller2011product}, there is a direct influence of the covariates on the definition of the partition distribution. This representation remains sample size consistent. 

The predictive model follows a similar approach: when considering a new observation $(n+1)$, it is assigned to either a new cluster or one of the existing clusters. The assignment probabilities are proportional to the marginal likelihoods evaluated at the covariate value of the new observation. These probabilities can vary across clusters, indicating that if the covariate value of observation $n+1$ is close to the value $x_h^*$ of a subject in cluster $h$, then the subject will be allocated with a higher probability to cluster $h$.

Similar to \cite{muller2011product}, the covariates $x_i$ are assumed to follow an auxiliary model, which can be chosen to be conjugate for computational efficiency.

\subsubsection{Spatial extensions.} In a spatial setting, \cite{page2016spatial} propose a flexible location-dependent product partition model that incorporates spatial information by considering the likelihood of assigning locations that are far apart to the same cluster. Let $s_1, \ldots, s_n$ denote $n$ locations, which can be either two-dimensional coordinates or areal locations. To extend the product partition model to include spatial information, the cohesion function can be modified as follows:
$$
p(\rho_n | s^*_1, \ldots, s^*_{K}) \propto \prod_{h=1}^{K} c(S_h, s_h^*).
$$
where $s^*_h = \{s_i: i \in S_h\}$. One possible approach is to define the cohesion function similar to the one used in the DP: 
$$
c(S,s^*) = \begin{cases}
\alpha \times \Gamma(|S|) & \mbox{if } S \mbox{ is spatially connected,} \\
0 & \mbox{otherwise}.
\end{cases}
$$
However, this model is computationally challenging to approximate. \cite{page2016spatial} propose four alternative functions that define the cohesion as a decreasing function of the distance between locations. Some of these functions exhibit sample size consistency, while others do not. Unlike the DP, \cite{page2016spatial} do not derive an analytic formula for the expected number of clusters in the spatial extension of the product partition models because the expectations depend on the distances among locations. However, they provide experimental results demonstrating that the expected number of clusters may grow at a slower or faster rate compared to a standard DP, depending on the specific cohesion function chosen.

\subsection{Alternatives to product partition models}

Alternatively to product partition models or induced partitions, it is possible to define other distributions for the random partition. There are several possible prior distributions to consider. 

\subsubsection{Uniform prior.} The simplest one is the uniform prior:
$$
p(\rho_n = (S_1, \ldots, S_{K})) = \frac{1}{\mathcal{B}_n}
$$
where $\mathcal{B}_n$ is the Bell number.


\subsubsection{Hierarchical prior.} \cite{casella2014cluster} propose a hierarchical uniform prior where the prior distribution on the partitions is conditioned on the number of components, which influences the number of clusters. The prior distribution is defined as:
$$
p(K,C_1, \ldots, C_n) = p(C_1, \ldots, C_n| K) p(K).
$$
The prior distribution on $(C_1, \ldots, C_n)$ can be chosen to be uniform, while $p(K)$ can be chosen in such a way that it assigns small support to the case where $K=n$.

\subsubsection{Ewens-Pitman prior.} \cite{dahl2017random} propose an Ewens-Pitman attraction (EPA) distribution, which allocates observations based on their ``attraction'' to existing clusters. The attraction to a given cluster is determined by pairwise similarities between the current observation and the observations already in the cluster. The allocation process sequentially assigns items to subsets, creating a partition. Let $(j_{(1)}, \ldots, j_{(n)})$ be permutations of ${1,\ldots,n}$ such that the $i$-th allocated observation is $j_i$. The resulting partition of $n$ observations has $K^{(i)}$ subsets. To make allocation decisions, a similarity function $\lambda(j_{(i)},j_{(\ell)})$ is required. It is common to define the similarity function as a function of the distance between observations, i.e., $\lambda(j_{(i)},j_{(\ell)}) = f(d_{i\ell})$, where $f$ is a non-increasing function. As the function $f(d) \rightarrow 0$, the EPA distribution becomes increasingly different from the Ewens distribution, which is the partition distribution of a DP. This means that the EPA distribution favors partitions that group items with small distances, contrasting the behavior of the DP. 

The EPA distribution is defined as the product of increasing conditional probabilities:
$$
p(\rho_n = (S_1, \ldots, S_K))| \alpha, \delta, \lambda, (j_{(1)}, \ldots, j_{(n)})) = \prod_{i=1}^n p_i(\alpha,\delta, \lambda, (j_{(1)}, \ldots, j_{(n)})),
$$
where
$$
p_i(\alpha,\delta, \lambda, (j_{(1)}, \ldots, j_{(n)})) =\begin{cases}
\frac{i-1-\delta K^{(i-1)}}{\alpha+i-1} \frac{\sum_{j_{\ell} \in S} \lambda(j_i, j_{\ell})}{\sum_{\ell=1}^{i-1}\lambda(j_i, j_{\ell})} & S\in \rho_n^{(i-1)}, \\
\frac{\alpha + \delta K^{(i-1)}}{\alpha + i-1} & S \text{ being a new subset}.
\end{cases}
$$
Here, $S$ represents the cluster to which observation $i$ is allocated, and $\rho_n^{(i-1)}$ is the partition of the first $(i-1)$ observations. The ratio of the similarity functions provides the ``attraction'' of $j_i$ to the observations allocated to $S$. The distribution is invariant to permutations of the observations and also to scale changes in the similarity function $\lambda$. The parameter $\delta \in [0,1)$ is a discount parameter, while $\alpha$ is a concentration parameter.

The EPA distribution also produces a probability distribution on the number of clusters $K_+$, which can be derived in closed form. This distribution does not depend on the similarity $\lambda(j_{(i)}, j_{(\ell)})$ or the permutation $(j_{(1)}, \ldots, j_{(n)})$. The expected number of clusters is $\mathbb{E}[K_+] = \sum_{i=1}^n w_i$, where $w_1 = 1$ and $w_i = \frac{\alpha + \delta \sum_{\ell=1}^{i-1} w_{\ell}}{\alpha+i-1}$ for $i>1$. As $n$ increases, the average number of clusters increases, with the rate of growth depending on $\alpha$ and $\delta$. In particular, as $\alpha$ and $\delta$ approach zero, the average number of clusters increases more slowly.

The PPF of the EPA distribution is given by:
$$
p_j(n_1, \ldots, n_{K}) = \left( \frac{n_h-\delta K n_h}{n+\alpha} \right) \mathbb{I}\{1 \leq h \leq K\}  + \left( \frac{\alpha + \delta K}{n+\alpha}\right) \mathbb{I}\{h =K+1\}.
$$
For $\delta=0$ and $\lambda(j_{(i)},j_{(\ell)})$ constant for all $i,\ell$, this PPF corresponds to the PPF of a DP, i.e., the Ewens distribution; see Equation \eqref{eq:PPFEwens}. However, there is no way to recover the PPF of the Ewens-Pitman distribution; see Equation \eqref{eq:PPFEP}. The EPA distribution applies the discount $\delta$ proportionally to the relative size of the cluster and the number of clusters, whereas the Ewens-Pitman distribution applies a uniform discount $\gamma_a$ to small and large clusters. This difference increases the entropy of the derived partitions for the EPA distribution and decreases the proportion of singletons. However, the distribution on the number of clusters is the same for both distributions.

Similarly to product partition models, the EPA distribution is symmetric. However, unlike product partition models, it is not marginally invariant. Nevertheless, in product partition models, the hyperparameters are often fixed to constant values for computational feasibility. On the other hand, the EPA model can easily treat the hyperparameters as random variables and estimate them from the data points. Furthermore, a characteristic of the EPA distribution is that it allocates probability among partitions within a given number of clusters, but it does not redistribute probability among sets of partitions with different numbers of clusters.

\subsubsection{Informative prior distributions.} None of these prior distributions allows for the incorporation of prior information about the grouping. \cite{paganin2021centered} propose an approach where it is possible to incorporate prior information on the partition. The prior for $S = (S_1, \ldots, S_{K})$ can be defined as:
$$
p(S | S^{(0)}) \propto p_0(S^{(0)}) e^{- d(S, S^{(0)}; \psi)},
$$
where $S^{(0)}=(S_1^{(0)}, \ldots, S_1^{(0)})$ and $\psi$ is a penalisation parameter. As $\psi$ increases, the model favors partitions $S$ that are similar to $S^{(0)}$, but not in a uniform way, as the space of partitions $\mathcal{P}_n$ is not uniform. In other words, for a fixed configuration, there is a heterogeneous number of partitions. The function $d(\cdot, \cdot)$ is a distance measure on the partitions, such as the Variation of Information by \cite{meilua2007comparing}, and $p_0(S^{(0)})$ is a baseline EPPF. 
The baseline EPPF can be chosen as the uniform EPPF:
$$
p_0(S^{(0)}) = \frac{1}{\mathcal{B}_n}
$$
where $\mathcal{B}_n$ is the Bell number. The concentration around $S^{(0)}$ depends on $n$.

The distance $d(S, S^{(0)})$ takes a finite number of discrete values ${\tau_1, \ldots, \tau_L}$, where $L$ depends on $S^{(0)}$ and $d(\cdot, \cdot)$. Let $S_{\ell}(S^{(0)}) = {S \in \mathcal{P}n : d(S, S^{(0)}) = \tau{\ell}}$ be the set of partitions having the same fixed distance from $S^{(0)}$. The exponential term penalizes partitions in the same set $S_{\ell}(S^{(0)})$ equally for a given $\tau_{\ell}$, but the resulting probability may differ depending on the baseline EPPF $p_0(S^{(0)})$.

\section{Clustering populations}
\label{sec:clust_subpop}

Several hierarchical models have been introduced based on the DP to cluster observations at multiple levels. To address the clustering of populations, it is beneficial to introduce the concept of partial exchangeability. A set of random variables $(Y_{1,1}, Y_{1,2}, \ldots, ,Y_{2,1}, Y_{2,2}, \ldots)$ is considered partially exchangeable if, for all sample sizes $n_1, n_2 \geq 1$, and all permutations $(i_{(1)}, \ldots, i_{(n_1)})$ and $(j_{(1)}, \ldots, j_{(n_2)})$ of $(1,2,\ldots,n_1)$ and $(1,2,\ldots,n_2)$, the distribution
$$
f(y_{1,1}, \ldots, y_{1,n_1}, y_{2,1}, \ldots, y_{2,n_2}) = 
	f(y_{1,i_{(1)}}, \ldots, y_{1,i_{(n_1)}}, y_{2,j_{(1)}}, \ldots, y_{2,j_{(n_2)}})
$$
remains the same. In other words, the distribution of joint samples remains invariant under permutations within each sample. The concept of partial exchangeability can be extended by considering random variables $Y_{ij}$, where $j=1,\ldots, J$ and $i=1, \ldots, n_j$, and each $Y_{ij} \sim f_j$, with $f_j \sim H$ repesenting a prior on the space of random measures. The entire sequence of random variables is exchangeable if the probability measure $H$ assigns probability one to $\{(f_1, \ldots, f_J) \in \mathcal{F}^J_{\mathcal{Y}}: f_1 = f_2 = \ldots = f_J \}$, for some class of distributions $\mathcal{F}^J_{\mathcal{Y}}$. The opposite of exchangeability is independence. However, in practical situations, it can be useful to consider intermediate scenarios where random measures are similar but not exactly identical. The situation of partial exchangeability is associated with an allocation distribution described by a partial EPPF (pEPPF).

\subsection{Different levels of clustering}
\label{sub:popintro}

\subsubsection{Focus on the first level of clustering.} \cite{reich2011spatial} utilize a separable structure to identify genomic clusters and their relationship to spatial locations in order to investigate the interplay between natural selection and environmental factors. In their model, the allele frequencies for individual $i$ at locus $\ell$ follow a multinomial distribution, conditioned on the cluster assignment:
\begin{equation*}
    Y_{i\ell} | C_i = c_i \sim \mathcal{M}\mbox{ultinomial}(2, \omega_{c_i \ell})
\end{equation*}
where $\omega_{c_i \ell}$ represents the vector of allele probabilities at locus $\ell$ in cluster $c_i$. These probabilities are assigned a stick-breaking prior. Separately, the spatial locations are modeled nonparametrically, conditioned on the cluster assignment:
\begin{align*}
    s_i | C_i = c_i &\sim F_{c_i} \\
    f_{c_i}(s_i) &= \sum_{h=1}^{\infty} \pi_{c_i h} \mathcal{N}(s_i | \mu_{c_i h}, \Sigma_{c_i}),
\end{align*}
where $\mathcal{N}(\mu, \Sigma)$ denotes a multivariate normal distribution with mean vector $\mu$ and covariance matrix $\Sigma$. Lastly, the cluster assignment is modeled as a categorical variable:
\begin{equation*}
C_i \sim \mathcal{C}\text{at}(q_1, \ldots, q_K),
\end{equation*}
where $K$ could potentially be infinite and the probabilities $q_1, \ldots, q_K$ are assigned a stick-breaking construction. In the study by \cite{reich2011spatial}, there are multiple clustering features at each step, but the primary focus lies in the clustering of spatial locations, characterized by the allocation variable $(C_1, \ldots, C_n)$.

\subsubsection{Clustering for meta-analysis.} \cite{muller2004hierarchical} introduce a hierarchical approach for meta-analysis problems. The random distribution of observations is defined as a mixture of a common measure $F_0$, representing the shared component across all populations, and a random measure $F_j$ specific to population $j$. In this framework, the model can be described as follows:
\begin{align}
    Y_{1,j}, \ldots, Y_{n_j,j} &\sim  f(y_{1,j}, \ldots, y_{n_j,j} | \theta_{ij})  \qquad j=1, \ldots, J\nonumber \\
    \theta_{ij} &\sim H_j \qquad i=1, \ldots, n_j, \quad j=1, \ldots, J \nonumber \\ 
    H_j &= \gamma F_0 + (1-\gamma) F_j \label{eq:hiearch_mixt}\\
    F_j &= \text{discrete RPM}, \nonumber
\end{align}
where $0 \leq \gamma \leq 1$ represents the weight determining the dependence among populations and the amount of information borrowed by the estimation procedure from other probability measures. Thus, all data contribute to learning $F_0$, while $(y_{1,j}, \ldots, y_{n_j,j})$ contributes to the specific learning of $F_j$. \cite{kolossiatis2013bayesian} propose selecting $\gamma$ in a way that ensures $H_j$ is marginally a DP.

As a prior distribution for $\gamma$,  \cite{muller2004hierarchical} suggest:
\begin{equation*}
p(\gamma) = w_0 \delta_0(\gamma) + w_1 \delta_1(1-\gamma) + (1-w_0-w_1) Beta(\gamma | a_{\gamma}, b_{\gamma}),
\end{equation*}
where $w_0$ and $w_1$ assign non-zero probability to $\gamma = 0$, corresponding to independent $H_j$, and $\gamma = 1$, corresponding to exchangeable observations across populations. Additionally, $\gamma$ can follow a beta distribution with parameters $a_{\gamma}$ and $b_{\gamma}$ with positive probability. A similar model has been employed by Wang et al. (2019) to combine information from randomized and registry studies for causal inference. 
 
\subsubsection{Extensions of \cite{muller2004hierarchical}.} The work of  \cite{muller2004hierarchical} has been extended by \cite{dunson2006bayesian} which introduced latent trait distributions. \cite{caron2007bayesian} propose an alternative approach to incorporate temporal dependence using a generalized P\'olya urn, which represents time-varying DP mixtures. \cite{caron2014bayesian} utilize a mixture of DPs for heterogeneous ranking data with nonparametric Plackett-Luce components, where each component is parameterized by a random measure, such as a gamma process. \cite{billio2019bayesian} suggest employing model \eqref{eq:hiearch_mixt} for parameter blocks in high-dimensional vector autoregressive models.

A characteristic of model \eqref{eq:hiearch_mixt} is that the atoms are different for each population, even if they originate from the same component in the mixture. Conversely, \cite{gutierrez2019bayesian} propose an approach where the atoms for two populations can assume the same values, which is particularly useful when testing equality between two or more random measures. Finally, \cite{lijoi2014bayesian} propose a model where $F_0$ and $F_j$ are independent normalized completely random measures (NCRM). Although these two approaches may appear similar, the approach proposed by \cite{lijoi2014bayesian} cannot be interpreted as a generalization of the approach proposed by \cite{muller2004hierarchical}.  The main difference lies in the fact that the measures $H_j$ in \cite{muller2004hierarchical} are not guaranteed to be marginally DPs, while they are ensured to be marginally NCRM in the approach by \cite{lijoi2014bayesian}.

\subsection{Nested processes}

The models presented in Section \ref{sub:popintro} lack formal definitions of the induced partition model. One possible approach to defining partitions of populations is through nested models, which involve nesting discrete random probability measures. Nested DPs have been introduced by \cite{rodriguez2008nested} to perform both clustering among observations and clustering among distributions. For the case of $d=2$ populations, the model is defined as:
\begin{align*}
(Y_{i_1,1}, Y_{i_2,2}) | f_1, f_2 & \stackrel{ind}{\sim}f_1 \times f_2 \qquad i_1=1, \ldots, n_1, \quad i_2=1, \ldots, n_2 \\
f_1, f_2 | H & \stackrel{ind}{\sim} H \\
H &\stackrel{d}{=} \sum_{h=1}^{\infty} \pi_{h} \delta_{\theta^*_h} \\
\theta^*_h &\stackrel{i.i.d}{\sim} F = \sum_{\ell=1}^{\infty} w_{\ell} \delta_{\psi^*_{\ell}} \\
\psi^*_{\ell} &\stackrel{i.i.d}{\sim} F_0 = DP(\alpha, F_{00}).
\end{align*}
Here, for $h=1,2,\ldots$, $\pi_{h}$ are independent of $\theta^*_h$, and for $\ell=1,2,\ldots$, $w_{\ell}$ are independent of $\psi_{\ell}^*$. The extension to $J$ populations is straightforward. \cite{rodriguez2008nested} propose using DPs at both levels of the hierarchy, but other processes can also be employed. The advantage of using DPs is that the weights at both levels can be constructed using the stick-breaking representation, which offers computational efficiency. As $H$ is almost surely discrete, $f_1$ and $f_2$ can be equal with positive probability, implying that $Y_{i_1,1}$ and $Y_{i_2,2}$ can have the same distribution.

\cite{camerlenghi2019latent} demonstrate that nested DPs are unable to flexibly and realistically cluster populations. Specifically, if $f_1$ and $f_2$ have at least one common atom, the posterior distribution of $(f_1, f_2)$ degenerates to the case where $f_1 \stackrel{d}{=} f_2$. This characteristic, where clusters can be either entirely common among populations or entirely distinct, is not unique to nested DPs but is present in all nested processes.

To address this degeneracy issue, \cite{camerlenghi2019latent} introduce latent nested processes where the nesting structure is applied to the underlying completely random measures. This allows for a representation in which each distribution $f_j$ can be expressed as a mixture:
$$
f_{j}  = \frac{\mu_{j} + \mu_{S}}{\mu_{j}(\mathcal{Y}) + \mu_S(\mathcal{Y})} = \gamma_j \frac{\mu_{j}}{\mu_{j}(\mathcal{Y})} + (1-\gamma_{j}) \frac{\mu_S}{\mu_S(\mathcal{Y})} \qquad j=1,2, \ldots, J.
$$
Here, $\mu_1, \mu_2, \ldots, \mu_J, \mu_S$ are normalized random measures with independent increments, and $\mathcal{Y}$ represents the sample space. This representation allows $f_{j}$ to be a mixture of a population-specific component $\mu_{j}$ and a common component $\mu_S$. As a result, two distributions, $f_1$ and $f_2$, can share some related atoms (and clusters) while also having distinct clusters specific to each population. The value of $\gamma_j$ determines the degree of relatedness between populations: when $\gamma_j=1$ for a specific population $j$, the populations are independent; when $\gamma_j=0$ for all populations, the populations are exchangeable. Furthermore, the latent nested process can represent all intermediate situations between independence and full exchangeability. To test equality among populations, one can examine the posterior distribution of $\mathbb{I}[\mu_j=\mu_\ell]$ for $j \neq \ell$. Additionally, the latent nested process induces a pEPPF, which is a linear combination of the EPPF corresponding to the fully exchangeable case and the EPPF corresponding to unconditional independence.

It is possible to incorporate a dispersion parameter, either scalar or infinite-dimensional, that governs the variability among samples within the same populations and can be assigned a hyperprior. In this case, the completely random measure can be defined as $\frac{\mu_j}{\mu_j(\mathcal{Y}) \times \Omega}$, where $\Omega$ represents the space over which the dispersion parameter is defined \citep{christensen2020bayesian}.

The main drawback of this model is its computational cost. While the model allows for a latent representation associated with the allocation to each cluster, each step of the corresponding MCMC requires approximating integrals, which can be computationally demanding when using Monte Carlo integration. This slows down the estimation procedure and makes generalizing the model to the case of $J>2$ populations infeasible in realistic situations.

An alternative approach is to select $\mu_{j}$ and $\mu_S$ as independent gamma processes, with $\mu_{j}$ being independent and identically distributed. In this case, $\frac{\mu_j}{\mu_j(\mathcal{Y})}$ and $\frac{\mu_S}{\mu_S(\mathcal{Y})}$ are draws from two independent DPs, and the resulting process is a latent nested DP \citep{beraha2021semi}. Another option is to use a PY process instead of a DP, which introduces more flexibility and allows for extending the model in the presence of covariates. For example, $\mu_j$ and $\mu_S$ can be defined as gamma processes in $\mathcal{Y} \times \mathcal{X}$, where $\mathcal{X}$ represents the covariate space.

Another potential drawback of the latent nested process is that $\mu_S$ includes all the common atoms across populations, but these common atoms must have the same weight across populations. This implies that different distributions sharing the same clusters should also have observations allocated to those clusters in the same proportions. This constraint can be limiting in certain scenarios.

To address this limitation, one possible solution is to introduce a weight matrix that relates the weights to a matrix of indicators, where each row represents a population. While this approach is straightforward to implement, it results in a more complex mathematical model and lacks a closed-form expression for the pEPPF. For further information, see \cite{liu2019invited} and \cite{soriano2019mixture}, which discuss these issues in the context of latent nested processes.

Nested processes have been introduced, for example, for analysing differences in quality of care across hospitals and states \citep{rodriguez2008nested}, where states represent the higher level of clustering.

\subsection{Hierarchical processes}

Unlike the approaches taken by \cite{muller2004hierarchical} and \cite{rodriguez2008nested}, \cite{teh2005sharing} propose a different model that incorporates information sharing among observations through a common prior. In this model, the mean and covariance of each component are shared across all samples, while the mixture weights remain unique. Each random measure $F_j$ is distributed according to a specific $DP(\alpha_{0}, F_{0})$, where the base measure $F_0$ has a DP prior, denoted as $DP(\gamma, F_{00})$. The clustering in this model arises from the shared clusters among groups of observations.

To elaborate further, the hierarchical DP (HDP) is a nonparametric prior process in which observations $Y_{ij}$, for $i=1,\ldots, n_j$ and $j=1, \ldots, J$, are distributed according to a generic distribution $f(\theta_{ij})$, where $\theta_{ij}$ follows the distribution $F_j$ with a Dirichlet process prior. Each $F_j$ is conditionally independent given the base measure $F_0$, and $F_0$ itself follows a DP prior. The model can be summarized as follows:
\begin{align*}
    Y_{ij} | \theta_{ij} &\stackrel{ind}{\sim} f(\theta_{ij}) \qquad i = 1, \ldots, n_j \quad j=1, \ldots, J \\
    \theta_{ij} | F_j &\stackrel{ind}{\sim} F_j \\
    F_j | \alpha, F_0 &\sim DP(\alpha,F_0) \\
    F_0 | \gamma, F_{00} &\sim DP(\gamma, F_{00}).
\end{align*}
This hierarchical model allows individual $F_j$ to share atoms. This sharing is evident from the stick-breaking construction:
\begin{align*}
    F_0 = \sum_{h=1}^{\infty} w_h \delta_{\theta^*_h} \qquad \mbox{and} \qquad F_j = \sum_{h=1}^{\infty} \pi_h \delta_{\theta^*_h}.
\end{align*}
Thus, the model for each $j$ relies on groups sharing the same mixture atoms $\theta^*_h$, but with different mixing weights $(\pi_1, \pi_2, \ldots)$. The hierarchical construction of the HDP allows the definition of clusters at different levels. Recently, \cite{camerlenghi2019distribution} characterize the posterior distribution of this prior process.

The distinction between the HDP and the nested DP and its extensions lies in their clustering properties. The nested DP constructs clusters of individuals across different groups, where the random measures either share the same atoms and weights or have no sharing. In contrast, the HDP allows random measures to share the same atoms but with different weights. As a result, the nested DP enables clustering at both the level of observations and the level of distributions, while the HDP only facilitates clustering at the level of observations.

In order to apply the HDP for population clustering, \cite{beraha2021semi} introduce the semi-HDP, which incorporates a baseline distribution as a mixture between a DP and a non-atomic measure. This construction reduces the computational burden compared to \cite{camerlenghi2019latent} for dimensions $J>2$, and allows for population clustering through a random partition model. The model can be represented as follows:
\begin{align*}
Y_{ij} | F_1, \ldots, F_J, C_1, \ldots, C_J& \overset{ind}{\sim} \int_{\Theta} f(\cdot | \theta) F_{c_j}(d\theta)\qquad i=1, \ldots, n_j, \quad j=1, \ldots,J \\
C_1, \ldots, C_J & \sim Cat(\pi_1, \ldots, \pi_J) \\
F_1, \ldots, F_J | F_0 &\sim DP(\alpha, F_0) \\
F_0 & = \gamma G_0 + (1-\gamma) G \\
G &\sim DP(\kappa, G_{00}) \\
\gamma &\sim Be(a_{\gamma},b_{\gamma}), 
\end{align*}
where $Cat(\pi_1, \ldots, \pi_J)$ is a categorical distributions, with weights $\pi_1, \ldots, \pi_J$, that means that there are at most $J$ populations; the vector $\pi_1, \ldots, \pi_J$ can be assigned a (Dirichlet) prior distribution. $F_j$ is a discrete random probability measure, i.e. $F_j = \sum_{h =1}^{\infty} w_{jh} \delta_{\theta^*_{jh}}$ where $w_{jh}$ are given a stick-breaking construction and $\theta^*_{jh} \sim F_0$. $F_0$ is a mixture between a DP $G$ and a fixed probability measure $G_0$. $G = \sum_{h =1}^{\infty} w_h \delta_{\psi_h}$, where $w_h$ are given a stick-breaking construction and $\psi_h \sim G_{00}$; $\alpha$ and $\kappa$ are two positive concentration parameters. 

When $\gamma=1$, all atoms and weights in $F_j$ are independent and distinct. When $\gamma=0$, the semi-HDP reduces to the HDP model proposed by \cite{teh2005sharing}, where all $F_j$ share the same atoms, but with different weights. In contrast to \cite{rodriguez2008nested}, the semi-HDP allows for a positive probability of atom sharing across different $F_j$'s, and atoms can also be shared within the same $F_j$ since all $F_j$'s share the same atoms as $G$. Moreover, $F_j \neq F_{\ell}$, for $j \neq \ell$ with probability one because the weights are different, even when $0 < \gamma < 1$.

\cite{beraha2021semi} demonstrate that this prior exhibits full weak support and that the covariance between $F_j$ and $F_{\ell}$, for all $j, \ell$ in the set ${1,2,\ldots,J}$, depends on two parameters: the concentration parameter of the second level $\kappa$, and the mixing weight of the first level $\gamma$. As $\gamma$ approaches 1, $F_j$ and $F_{\ell}$ become increasingly uncorrelated. Additionally, \cite{beraha2021semi} derive the pEPPF, which is a convex combination of the EPPF corresponding to the fully exchangeable case and the product of the marginal EPPFs for each $j$.

The HDP has been widely successful and applied in various fields. It has been used in cytometry \citep{cron2013hierarchical}, genomics \citep{sohn2009hierarchical}, social networks \citep{airoldi2008mixed}, imaging \citep{sivic2005discovering}, health sciences \citep{gaba20202}, topic models \citep{gerlach2018network}, neuroimaging \citep{jbabdi2009multiple, wang2011tractography}, visual scenes \citep{kivinen2007learning}, and many other domains.

The motivating example of \cite{teh2005sharing} comes from information retrieval and involves modeling relationships among documents. The main goal is to allow topics to be shared among documents and to enable groupings of documents. At the higher level, each document is a mixture of topics, which can be clustered, and at the lower level, each document consists of a bag of words. \cite{beraha2021semi} apply the semi-HDP to compare the effectiveness of lecturers teaching the same course to different cohorts. They focus primarily on clustering at the higher level (the lecturers) and producing accurate density estimates of the lower levels (students' marks).

\section{Posterior distributions on the partitions}
\label{sec:posteriors}

A model with a prior distribution on the partition $\rho_n$ gives rise to a posterior distribution on the partition. However, estimating the posterior partition in the context of clustering is challenging, since it is typically a high-dimensional problem and it is unlikely that the MCMC algorithm visits a specific partition more than once, therefore several computational solutions have been proposed in the literature to obtain posterior estimates of the partition. 

\subsection{Stochastic search methods.} 

Stochastic search methods have been employed to estimate the posterior mode of $\rho_n$ at the end of the MCMC algorithm, as discussed in \cite{brunner1999bayesian} and \cite{nobile2007bayesian}. To approximate the posterior mode of the partition, \cite{dubey2003clustering}, \cite{heller2005bayesian}, and \cite{heard2006quantitative} utilize Bayesian deterministic hierarchical procedures, avoiding the need for MCMC sampling. Another approach, proposed by \cite{medvedovic2004bayesian}, involves hierarchical agglomerative clustering, where the distance is based on the posterior similarity matrix estimated through MCMC. While hierarchical clustering methods reduce the involvement of the experimenter in terms of the estimation procedure, they require a method for ``cutting the tree'', i.e., determining the optimal number of clusters. \cite{fritsch2009improved} propose a method to cut the tree by minimizing the Monte Carlo estimate of a posterior expected loss. This approach is implemented in the \texttt{mcclust} package in \texttt{R}.

\subsection{Decision-theoretic approaches.} 

\subsubsection{Maximum a posteriori.} From a decision-theoretic perspective, it is reasonable to seek the optimal partition that minimizes a posterior expected loss function:
$$
\rho^*_n = \arg \min_{\hat{\rho}_n} \mathbb{E}[L(\rho_n, \hat{\rho}_n)| y_1, \ldots, y_n] = \arg \min_{\hat{\rho}_n} \sum_{\rho_n \in \mathcal{P}_n} L(\rho_n, \hat{\rho}_n) p(\rho_n | y_1, \ldots, y_n),
$$
where $p(\rho_n | y_1, \ldots, y_n)$ is the posterior distribution of the partition $\rho_n$, and $L(\cdot)$ is a loss function. One possible approach is to select the Maximum a Posteriori (MAP) clustering, which corresponds to the optimal Bayesian estimate under the 0-1 loss function $L(\rho_n, \hat{\rho}_n) = \mathbb{I}[\rho_n \neq \hat{\rho}_n]$. However, the estimated MAP obtained from an MCMC output is unlikely to be representative, as a comprehensive exploration of the partition space is infeasible. Additionally, this loss function does not consider any notion of similarity between partitions.

\cite{dahl2009modal} proposes an algorithm to derive the MAP partition for a class of univariate product partition models. This algorithm guarantees finding the maximum a posteriori clustering or, at the very least, the maximum likelihood clustering when the partition model can be expressed in terms of a product partition distribution. The algorithm of \cite{dahl2009modal} can be applied when two conditions are met: the components in the modal clustering do not overlap (i.e., $S_h$ does not contain integers between the smallest and largest integers in $S_k$, or vice versa), and the cohesion function $c(S)$ depends only on the number of items contained in $S$. This algorithm can be highly efficient since it requires only $n(n+1)/2$ evaluations. However, it is restricted to univariate observations and does not provide an estimation error quantification. The algorithm is implemented in the \texttt{modalclust} package in \texttt{R}.

\subsubsection{Methods based on the Binder loss.} Posterior modes can become increasingly unrepresentative of the posterior distribution of the partition as the number of items increases. To address this issue, \cite{lau2007bayesian} propose using the Binder loss \citep{binder1978bayesian} instead of the 0-1 loss, aiming to respect the exchangeability in the labeling of clusters and items. The Binder loss penalizes pairs of items that are assigned to different clusters when they should be clustered together, and vice versa. This loss function is commonly used in Bayesian clustering because it can be expressed in terms of the posterior similarity matrix, which is an $n \times n$ matrix where the $(i,j)$-th element represents the posterior probability that observation $i$ is allocated together with observation $j$. The posterior similarity matrix can be easily estimated using MCMC.

In terms of allocation variables $C_i$ for $i \in {1,2,\ldots, n}$, the Binder loss is defined as:
$$
L(\rho_n, \hat{\rho}_n) = \sum_{(i,j) \in [n]} a \mathbb{I}_{[C_i = C_j, \hat{C}_i \neq \hat{C}_j]} + b \mathbb{I}_{[C_i \neq C_j, \hat{C}_i = \hat{C}_j]},
$$
where $a$ and $b$ are non-negative constants that represent the costs of pairwise misclassification. Specifically, $a$ represents the cost of not clustering observations that should be together, while $b$ represents the cost of clustering observations that should not be together. 
\cite{lau2007bayesian} propose an algorithm where optimizing the expected posterior Binder loss is formulated as a binary integer programming problem. In this formulation, the binary variable $X_{ij}=\mathbb{I}_{[\hat{C}_i = \hat{C}j]}$ is used, and the objective function (the posterior expected Binder loss) is a linear combination of variables $X_{ij}$ with weights $\Pr(C_i = C_j | y_1, \ldots, y_n)$. The algorithm iteratively targets the objective function for each item $i$, reassigning it optimally by either assigning it to an existing cluster or creating a new cluster. The algorithm leverages the fact that minimizing the posterior expectation of the Binder loss is equivalent to minimizing:
$$
\sum_{i \leq j} \mathbb{I}[\hat{C}_i = \hat{C}_j] \left( p_{ij} - \frac{b}{a+b}\right),
$$ 
where $p_{ij} = \Pr(C_i = C_j | y_1, \ldots, y_n)$ is the $(i,j)$-th element of the posterior similarity matrix. However, this algorithm may suffer from scalability issues and is only applied to the Binder loss, without generalisations to other loss functions. 

The Binder loss exhibits some asymmetry, preferring to split clusters rather than merge them. In practice, this asymmetry in the Binder loss may lead to the identification of extra-small clusters in the optimal partition, particularly at the boundary between clusters. This is typical when $a=b$, but it can be mitigated by imposing $a>b$, meaning a higher penalty for allocating observations to different clusters when they should be clustered together. 

\cite{dahl2006model} suggests a least squares clustering criterion which seeks the clustering that minimizes:
$$
\sum_{i=1}^n \sum_{j=1}^n (\mathbb{I}[\hat{C}_i = \hat{C}_j] - \hat{p}_{ij})^2.
$$
Minimizing this criterion is equivalent to minimizing the Monte Carlo estimate of the posterior expectation of the Binder loss when $a=b$. However, the method of \cite{dahl2006model} is limited to searching among the partitions visited during the MCMC algorithm.

\subsubsection{Methods based on the Variation of Information.} \cite{wade2018bayesian} propose an approach to summarize the posterior distribution of the clustering structure through both point estimates and credible sets. Unlike \cite{lau2007bayesian}, \cite{wade2018bayesian} propose using the Variation of Information (VI) loss function, as developed by \cite{meilua2007comparing}, and demonstrate through extensive simulation that the Binder loss and the VI loss can yield very different optimal partitions. The VI loss compares the information in two clusterings with the information shared between them: 
$$
VI(\rho_n, \hat{\rho}_n) = H(\rho_n) + H(\hat{\rho}_n) - 2I(\rho_n, \hat{\rho}_n),
$$
where $H(\rho_n)$ and $H(\hat{\rho}_n)$ represent the entropy of each of the two partitions, which measures the uncertainty in cluster allocation. $I(\rho_n, \hat{\rho}_n)$ represents the mutual information between the two partitions. Using the VI loss avoids the choice of $a$ and $b$ as for the Binder loss, however the VI loss does not have a representation in terms of a posterior similarity matrix, making it computationally more expensive to evaluate.

\cite{wade2018bayesian} propose a greedy search algorithm to explore the partition space, which utilizes an approximation. Minimizing the posterior expectation of the VI loss is equivalent to finding the optimum:
$$
\rho_n^* = \arg \min_{\hat{\rho}_n} \sum_{i=1}^n \log \left( \sum_{j =1}^n \mathbb{I} [\hat{C}_{i}= \hat{C}_j]\right) -2 \sum_{i=1}^n \mathbb{E} \left[ \log \left( \sum_{j =1}^n \mathbb{I}[C_{i}=C_j, \hat{C}_{i}=\hat{C}_j] \right)|y_1, \ldots, y_n\right].
$$
The expectation in the second term can be approximated using an MCMC output. However, evaluating this approximation is computationally costly, as it scales as $O(T n^2)$, where $T$ is the number of MCMC simulations, then considering many candidate $\hat{\rho}_n$ can be computationally prohibitive. Therefore, \cite{wade2018bayesian} propose using Jensen's inequality to swap the logarithm and the expectation, obtaining a lower bound on the expected loss that is more efficient to evaluate,
reducing the complexity of the algorithm to $O(n^2)$ for a given $\hat{\rho}_n$. While this approximation reduces computational complexity, its impact on the estimated optimal partition is not clear. Specifically, the properties of the VI loss function found in \cite{meilua2007comparing} are not guaranteed to hold when applying Jensen's inequality. Another drawback of the algorithm is its dependence on initialization, so it is advisable to run it multiple times starting from different initial partitions. The complexity of this algorithm is $O(\ell n^2)$, where $\ell$ defines the number of partitions to consider at each iteration. \cite{wade2018bayesian} suggest $\ell=n$. This method is implemented in the \texttt{R} package \texttt{mcclust.ext}. 
 
\subsubsection{Methods based on a generic loss.}
\cite{rastelli2018optimal} also rely on a decision-theoretic framework to derive the optimal partition. Unlike \cite{wade2018bayesian}, they propose an approach that does not depend on the posterior similarity matrix and does not involve any approximation. This method can be used with any loss function $L(\rho_n, \hat{\rho}_{n})$ that considers the two partitions through the counts $n_{hk}$, which denote the number of data points allocated to group $h$ in partition $\rho_n$ and to group $k$ in partition $\hat{\rho}_n$. Since the approach does not require the posterior similarity matrix, its computational complexity in terms of $n$ is reduced to linear order.

The method begins by randomly selecting a partition with small clusters and iteratively reassigning one item at a time to either an existing cluster or a new cluster, depending on the minimal Monte Carlo estimate of the expected loss. Similar to \cite{wade2018bayesian}, the algorithm requires multiple runs to obtain a partition that is closer to the optimal solution. The approach also requires defining a maximum number of clusters $K_d$, which \cite{rastelli2018optimal} suggest setting equal to $n$, although this choice can increase complexity. The complexity of this algorithm is $O(T \cdot K_d^2 \cdot n)$; if $K_d = n$, the complexity becomes $O(T n^3)$. This method is implemented in the \texttt{R} package \texttt{GreedyEPL}, which provides support for various loss functions, including the Binder loss and the VI loss.

\subsubsection{Methods based on the generalised VI loss.} \cite{dahl2021search} provide a generalization of the original VI loss, similar to the original Binder loss, where weights $a$ and $b$ represent the cost of failing to cluster two observations that should be clustered together and clustering two observations that should not be clustered together, respectively. This generalization maintains the properties of the original VI loss and can be evaluated without incurring higher computational costs. The generalized VI loss with positive weights $a$ and $b$ is given by:
$$
L(\rho_n, \hat{\rho}_n) = a \sum_{S \in \rho_n} \frac{|S|}{2} \log_2 \frac{|S|}{2} + b \sum_{S^{\prime} \in \hat{\rho}_n} \frac{|S^{\prime}|}{2} \log_2 \frac{|S^{\prime}|}{2} - (a+b) \sum_{S \in \rho_n} \sum_{S^{\prime} \in \hat{\rho}_n} \frac{|S \cap S^{\prime}|}{n} \log_2 \frac{|S \cap S^{\prime}|}{n}.
$$
Here, $S=(S_1, \ldots, S_K)$ and $S^{\prime} = (S^{\prime}_1,\ldots, S^{\prime}_{K^{\prime}})$ represent two partitions. This loss function can be targeted again by a greedy stochastic search algorithm.

The algorithm begins with an initialization step, either random or sequential. Then, one-at-a-time reallocation of individual observations is performed in a random order. Each observation is removed from its cluster and reallocated to either existing clusters or a new cluster, based on the choice that maximizes the Monte Carlo estimate of the posterior expected loss. This process is repeated until there is no change after a complete run on all $n$ observations. To avoid getting stuck in a local minimum, occasionally a cluster is ``killed'' by removing all observations from it and reallocating them sequentially to other clusters. If the Monte Carlo estimate of the expected loss at the end of this reallocation is not lower than the one obtained before the cluster was destroyed, the step is forgotten. This algorithm reduces to that of \cite{rastelli2018optimal} when the initialization is not sequential and the number of times clusters are destroyed is set to zero.

The complexity of this algorithm is $O(T \cdot K_d \cdot K_{MCMC} \cdot n)$, where $K_{MCMC}$ is the maximum number of clusters observed among the MCMC samples. This algorithm is implemented in the \texttt{R} package \texttt{salso}.

\subsubsection{Credible balls for the partition.} In addition to estimating the optimal partition, \cite{wade2018bayesian} also define credible balls:
$$
B_{\varepsilon^*}(\rho_n^*) = \{\rho_n: d(\rho_n^*, \rho) \geq \varepsilon^*\},
$$
where $\varepsilon^*$ is the smallest $\varepsilon > 0$ such that $\Pr(B_{\varepsilon}(\rho_n^* | y_1, \ldots, y_n)) \geq 1-\alpha$. The bounds of these balls are represented by partitions such that:
$$
\Pr(B_{\varepsilon^*}(\rho_n^*) | y_1, \ldots,y_n) = \mathbb{E}[\mathbb{I}[d(\rho_n^*,\rho_n) \leq \varepsilon]|y_1, \ldots, y_n],
$$
where the expected value can be estimated based on the partitions visited by the MCMC algorithm with positive probability. However, this definition of credible balls does not guarantee that all partitions inside the credible balls have a higher posterior probability than partitions outside the credible balls. Alternatively, one can consider the highest posterior regions and list all the partitions that have a posterior probability above a certain threshold.

\subsubsection{Methods uncertainty quantification for the partition.}

Following \cite{Escobar1995}, \cite{lavigne2024quantifying} derive the posterior predictive distribution for a DP mixture model as a finite mixture linked to the preferred partition identified by postprocessing of an MCMC output. The finite mixture is a convex conbination of $K_+$ clusters, where weights are the number of observations each cluster has been allocated. Such representations allows for an uncertainty quantification of the estimated partition.

\section{Conclusions}
\label{sec:conclu}

Clustering is a fundamental problem in statistics. Model-based clustering offers the advantage of introducing a probabilistic allocation of each observation to possible clusters, as well as a probabilistic definition of the number of clusters. 

In a Bayesian framework, a popular approach for model-based clustering is to impose a mixture model with an unknown number of clusters, using either a finite or infinite number of components. This leads to a random model on the partition of observations. Various models have been introduced, but the Dirichlet process mixture model and its extensions have been shown to be inconsistent for estimating the number of clusters, although some recent works seem to be optimistic and suggest potential improvements. Additionally, the prior distribution chosen for the number of components in a finite mixture model strongly affects the estimation of the number of clusters.

An alternative approach is to directly choose a model for the partition, such as through product partition models. While these methods aim directly at modeling clustering, they rely on assumptions about the partitions that may not always hold and can be difficult to verify.

In recent years, there has been attention towards an interesting extension of clustering, which is the problem of clustering populations. In this case, we have reviewed methods based on Gibbs-type priors that extend mixture models.

Finally, an interesting development is focused on the best ways to summarise the posterior distribution of the partition, and several decision-theoretic approaches have been compared. 

This work provides a review of the results and models proposed for Bayesian clustering. Some of the reviewed models were not necessarily introduced for clustering, particularly models that include covariates. However, they are frequently used in applied settings for clustering purposes. This work contributes to the existing literature by comparing the properties and limitations of available models for Bayesian clustering, taking into account the inconsistency result reported by \cite{miller2014inconsistency}. The proposed comparison aims to highlight the advantages and disadvantages of different methodologies, with the hope of inspiring new avenues for future research.

An interesting line of future research concerns clustering in high dimensions. It is well known that posterior inference in this context may suggest either a large number of clusters or, in some cases, a limited number of clusters. \cite{chandra2023escaping} provides conditions under which the posterior distribution of a random partition based on a finite sample tends to exhibit degenerate behavior, resulting in either all clusters being occupied or only one cluster being occupied. The solution proposed in \cite{chandra2023escaping} involves applying dimensionality reduction through a latent factor model and shrinkage priors. The findings presented in this review suggest that a prior on the partition may have a parsimonious effect on estimating the number of clusters and can guide future research directions. Works such as \cite{grazian2020loss} and \cite{page2023informed} suggest that building prior distributions using a decision-theoretic approach may improve the behavior of the estimation procedure for the partition, or help in understanding how informative the data are about the specific partition.

It is worth reflecting on the fundamental nature and significance of this problem. The number of clusters is inherently tied to the specification of the mixture components, which are often merely building blocks for a flexible density without any physical meaning. This raises the question: what truly constitutes the number of clusters, and is it even a meaningful or interesting quantity to estimate? Additionally, even when clusters have physical meaning, they might be misspecified, further complicating the notion of a ``true'' number of clusters. At a more basic level, the problem of Bayesian estimation of the number of sets in a random partition is not yet fully understood and solved. These reflections suggest that while the topic is important, it also warrants deeper theoretical and practical exploration.

\end{document}